\documentclass[twocolumn,twocolappendix]{aastex631}

\newcommand{\Msun}{\,M_{\odot}}

\usepackage{amsmath}
\shorttitle{Eruptive Mass Loss in Massive Stars}
\shortauthors{Cheng et al.}

\graphicspath{{./}{figures/}}

\begin{document}

\title{A Model for Eruptive Mass Loss in Massive Stars}

\correspondingauthor{Shelley J. Cheng}
\email{shelley.cheng@cfa.harvard.edu}

\author[0000-0002-5462-0294]{Shelley J. Cheng}
\affiliation{Center for Astrophysics $|$ Harvard \& Smithsonian, 60 Garden Street, Cambridge, MA 02138, USA}
\affiliation{Center for Computational Astrophysics, Flatiron Institute, 162 5th Avenue, New York, NY 10010, USA}

\author[0000-0003-1012-3031]{Jared A. Goldberg}
\affiliation{Center for Computational Astrophysics, Flatiron Institute, 162 5th Avenue, New York, NY 10010, USA}

\author[0000-0002-8171-8596]{Matteo Cantiello}
\affiliation{Center for Computational Astrophysics, Flatiron Institute, 162 5th Avenue, New York, NY 10010, USA}
\affiliation{Department of Astrophysical Sciences, Princeton University, Princeton, NJ 08544, USA}

\author[0000-0002-4791-6724]{Evan B. Bauer}
\affiliation{Center for Astrophysics $|$ Harvard \& Smithsonian, 60 Garden Street, Cambridge, MA 02138, USA}

\author[0000-0002-6718-9472]{Mathieu Renzo}
\affiliation{Steward Observatory, University of Arizona, 933 N Cherry Avenue, Tucson, AZ 85721, USA}
\affiliation{Center for Computational Astrophysics, Flatiron Institute, 162 5th Avenue, New York, NY 10010, USA}

\author[0000-0002-1590-8551]{Charlie Conroy}
\affiliation{Center for Astrophysics $|$ Harvard \& Smithsonian, 60 Garden Street, Cambridge, MA 02138, USA}

\begin{abstract}

Eruptive mass loss in massive stars is known to occur, but the mechanism(s) are not yet well-understood. One proposed physical explanation appeals to opacity-driven super-Eddington luminosities in stellar envelopes.  Here, we present a 1D model for eruptive mass loss and implement this model in the \texttt{MESA} stellar evolution code.  The model identifies regions in the star where the energy associated with a locally super-Eddington luminosity exceeds the binding energy of the overlaying envelope.  The material above such regions is ejected from the star.  Stars with initial masses $10-100 \Msun$ at solar and SMC metallicities are evolved through core helium burning, with and without this new eruptive mass-loss scheme. We find that eruptive mass loss of up to $\sim10^{-2}~M_\odot \mathrm{yr}^{-1}$ can be driven by this mechanism, and occurs in a vertical band on the HR diagram between $3.5 \lesssim\log(T_\mathrm{eff}/\mathrm{K}) \lesssim 4.0$. This predicted eruptive mass loss prevents stars of initial masses $\gtrsim 20~M_\odot$ from evolving to become red supergiants, with the stars instead ending their lives as blue supergiants, and offers a possible explanation for the observed lack of red supergiants in that mass regime.

\end{abstract}


\section{Introduction} \label{sec:intro}
Massive stars ($\gtrsim10\ M_\odot$) are important for understanding the evolution and structure of many astrophysical systems. When studying galaxies, massive stars are important since they impact the overall density and movement of matter which alters galaxy structure, dynamics and evolution \citep[e.g., see review by][]{Tacconi:20}. Additionally, mass loss from massive stars and their explosions play an important role in galaxy feedback \citep[e.g.,][]{Heckman:1990,Ceverino:09,Geen:15,Steinwandel:2023}. At smaller scales, quantifying the intrinsic mass loss from massive stars can help constrain the masses of compact objects \citep{Belczynski:10,Dominik:12,Renzo:17,Banerjee:20,Moriya:2022,Fragos:23}, study gravitational wave sources \citep{Ziosi:14,Mapelli:16,Uchida:2019,Vink:21}, and better understand the circumstellar environment responsible for observed signatures of interaction in early-time SN emission \citep[e.g.,][]{Schlegel:1990, Pastorello:2007, Morozova2017, Bruch2023, JacobsonGalan2024}. 

Observationally, massive stars are known to exhibit high levels of intrinsic mass loss \citep[e.g.,][]{Wood:83,Wood:92,Smith:2014,Meynet:2003}. In single massive stars, mass loss can generally occur in three ways; a radiation-driven wind, a pulsation-induced dust-driven wind, and through eruptions. One of the first analytical radiation-driven wind models was the CAK model \citep{CAK:75}, which predicted a mass loss rate on the order of $10^{-6}~M_\odot\ \text{yr}^{-1}$ for a O5-type main sequence star. The CAK model describes how the forces due to metal line absorption of photons can produce a mass outflow, and predicts mass loss rates and wind terminal velocities \citep[see, e.g.,][]{Lamers:1999, Owocki:1988, Smith:2014, cure:2023}. While the mass loss rate predictions of CAK were in general agreement with observations \citep{deJager:88,Nieuwenhuijzen:90,Nugis:2000}, the emissivity of the recombination processes that form emission lines depends on the square of density, so small density inhomogeneities or ``clumps" in the wind can cause an over-prediction of mass loss rates if they are not properly modeled. Subsequent work by \citet{Crowther:2002, Figer:2002, Hillier:2003, Massa:2003, Evans:2004, Puls:2006, Beasor:18, Beasor:20} have determined that CAK-based mass loss rates should be reduced by a factor of $2-10$ \citep{Crowther:2002,Hillier:2003,Fullerton:06,Puls:08,Smith:2014}. 

Pulsation-induced dust-driven winds are likely associated with red supergiants (RSGs) and AGB stars, where atmospheric shock waves caused by stellar pulsations -- and possibly convection -- are thought to lift gas to low enough temperatures to form dust which then collisionally couples with photons and accelerates outwards as a wind \citep[e.g.,][]{Liljegren:16, Willson:2000,vanLoon:05, Goldberg:22,Chiavassa:24}. Observations have suggested wind mass loss rates of up to $10^{-3}~M_\odot\ \text{yr}^{-1}$ in RSGs \citep{vanLoon:05}, but mass loss rates for RSGs are highly uncertain \citep{Smith:2014} and potentially overestimated \citep[e.g.,][]{Beasor:20,Beasor:2023,Decin:2024}. Many common prescriptions assume dust-enshrouded RSGs with higher mass loss rates \citep[e.g.,][]{Georgy:12,Ekstrom:12, Goldman:17}, and the choice of mass loss model can significantly alter the evolutionary  track of a RSG, with lower mass loss rates allowing the star to remain in the red but higher mass loss rates driving the star towards the blue \citep{Smith:2014, Renzo:17, Beasor:18,Beasor:20,Josiek:24} after spending time in the RSG phase. Observationally, RSGs exhibit an (apparent) maximum luminosity of $\log(L/L_\odot) \approx 5.5$, which corresponds to a maximum initial mass of around $20-30~M_\odot$ \citep{Massey:2000, Heger:2003, Levesque:2009, Ekstrom:12, DornWallenstein:23}. Post main-sequence mass loss has been suggested as a possible theory explaining this, particularly involving stars that exceed the Eddington luminosity limit \citep[e.g.,][]{Puls:08}. However, the specific nature of this mass loss is not well-understood, and the physical mechanism causing the dearth of RSGs above $20-30~M_\odot$ remains an open problem. Alternative explanations other than eruptive mass loss include envelope inflation \citep[e.g.,][]{Sanyal:15, Sanyal:17} and binary interactions \citep[e.g.,][]{Smith:2014, Morozova:2015, Margutti:2017}.

At even higher masses, luminous blue variables (LBVs) have shown very high mass loss rates of up to $0.1-5~M_\odot\ \text{yr}^{-1}$ \citep{Humphreys:1984,Smith:2017}. This high rate of mass loss seen in LBVs is observed to only occur for a short period of time, and as such is referred to as ``eruptive massive loss." Despite a few pioneering theoretical works \citep[e.g.,][]{Owocki:2004,Smith:2006}, efforts to form a comprehensive model for mass loss of massive stars have traditionally focused on developing scaling formulae based on empirical rates \citep[e.g.,][]{deJager:88,Nieuwenhuijzen:90,Vink:2001,vanLoon:05}. Additionally, \citet{Mauron:11} and \citet{Renzo:17} compare these scaling formulae for solar-metallicity non-rotating stars between $15-35~M_\odot$. However, these steady wind mass loss schemes do not attempt to and cannot reproduce the high rates of observed eruptive massive loss. Therefore, there has been some work that scales and tunes the theoretical wind mass-loss schemes using observations to reproduce observed rates of mass loss \citep[e.g.,][]{Ekstrom:12, Dessart:13, Meynet:15,Vink:23}. However, these schemes do not provide a physically motivated model that self-consistently captures the effect of eruptive mass loss.  

In this work, we begin with an overview of existing literature on radiation-pressure-dominated envelopes of massive stars from one-dimensional (1D) and three-dimensional (3D) models that motivates our work (Section \ref{sec:massive_stars}). We then develop an energy-based model for eruptive mass loss and implement it within the 1D stellar evolution software \texttt{MESA} (Section \ref{sec:methods}), and apply our model to massive stars of differing masses (between $10-100~M_\odot$) and metallicities (at $0.2~Z_\odot$ and $1~Z_\odot$). We explore the predicted rate of eruptive mass loss (Section \ref{sec:results}), and finally discuss our interpretation of our results and conclude in Section \ref{sec:discussion}.

\section{Radiation-pressure-dominated envelopes of massive stars} \label{sec:massive_stars}

Energy is transported in stars primarily through radiative diffusion and convection, including in the radiation-pressure-dominated extended outer regions of massive star envelopes, where both mechanisms are present. Due to opacity peaks in the envelope (from recombination of ions of iron, helium, and hydrogen), the local radiative luminosity may exceed the Eddington limit in layers of the envelope below sub-Eddington regions \citep{Jiang:15, Jiang:18}. Therefore, in these envelopes, density and gas pressure inversions may occur which are thought to contribute to outburst-like ``eruptive mass loss'' \citep{Joss:73, Humphreys:1994, paxton:13}. Specifically, the gas pressure increases outwards when radiative luminosity is super-Eddington (when $L_\mathrm{rad} > L_\mathrm{Edd}$), and under specific conditions, a density inversion may occur even if gas pressure is not inverted; see \citet{Joss:73} 
 and \citet{paxton:13} for more details. 

The envelopes of massive stars therefore present both physical and numerical challenges due to the presence of inversions and super-Eddington radiative luminosity. 1D numerical treatments can fail to find a hydrostatic equilibrium solution in these challenging regions \citep[e.g.,][]{paxton:13}. This is the case when using mixing-length-theory \citep[MLT, e.g.,][]{Bohm-Vitense58,Cox:68} or even time-dependent 1D formulations \citep[e.g.,][]{ Kuhfuss:86, Jermyn:2023}. Realistic 3D simulations show the limits of the 1D treatment, stressing the importance of the interplay between radiative transport and the complex landscape of turbulent fluctuations in the stellar envelope \citep{Jiang:15}. However these calculations are computationally very expensive, which has prevented a detailed study of the full parameter space relevant to massive stars. 

In the stellar evolution code \texttt{MESA} used in this work, several modifications to MLT aimed at these regions in the envelopes of massive stars are available, namely MLT++ \citep{paxton:13} and an implicit method \citep[\texttt{use\_superad\_reduction};][]{Jermyn:2023}. These methods involve artificially enhancing the efficiency of convection by reducing the superadiabaticity ($\nabla - \nabla_\mathrm{ad}$) in regions nearing the Eddington limit, which usually helps the code find a hydrostatic equilibrium solution. The first of these methods, MLT++, was introduced in \citet{paxton:13}. While MLT++ allowed evolution of massive stars to core-collapse, it lead to different efficiencies of energy transport in radiation-dominated stellar envelopes and the resulting evolution differs to that of convective energy transport by MLT. Subsequently, an implicit method was introduced in \citet{Jermyn:2023} that produces more local and robust results in closer agreement to that from MLT compared to MLT++ \citep[when comparing evolutionary tracks on the HR diagram from main sequence to RSG phase, see Figure~17 in Section~7.2 of][]{Jermyn:2023}. Nevertheless, while these 1D recipes can help improve numerical convergence and evolve massive stars further, they are not directly physically motivated.

Mass loss in massive stars has also been studied using 3D radiation-hydrodynamics (RHD) models. Indeed, 3D RHD models find enhanced convection in these radiation- and turbulent-pressure-dominated envelopes \citep[e.g.,][]{Jiang:15, Jiang:18,Schultz:23}. \citet{Jiang:15} and \citet{Jiang:18} simulated the envelope of massive stars at specific points in their evolution, and determined that non-steady mass loss outbursts potentially consistent with LBVs can be driven by helium and iron opacity peaks. Their predicted mass loss rates are significantly lower than those inferred from observations \citep{Humphreys:1984, Smith:2017}, possibly due to the fact that line-driving and dust formation are not included in the models. The importance of helium and iron opacity peaks in mediating continuum-opacity-driven mass-loss is not surprising, as the dependence of opacity on temperature and density near opacity
peaks can lead to radiation hydrodynamic instabilities \citep{Prendergast:73,Kiriakidis:1993, Blaes:2003, surez:2013}. 

As photons carrying energy diffuse outwards through the envelopes, they may encounter regions where photon diffusion occurs on a faster timescale than dynamical, convective, and acoustic processes (radiatively inefficient convection). This occurs at optical depths lower than the critical value $\tau_\mathrm{crit} \equiv c / c_\mathrm{s}$ \citep[e.g,][]{Jiang:15},
where $c$ is the speed of light and $c_\mathrm{s}$ is the sound speed. This can also be derived from balancing the diffusive radiation flux and the maximum convective flux in the radiation pressure-dominated
regime \citep[for more details see][]{Jiang:15}. This definition of critical optical depth is equivalent to the more general definition presented by \citet{Jermyn:22}  (see also discussions in Sections~2 and 4.4 of \citealt{Goldberg:22}).

The role of $\tau_\mathrm{crit}$ in determining convective regions where radiation nonetheless carries significant flux (often termed radiatively inefficient, or sometimes just ``inefficient" convection) has been demonstrated in 3D simulations. 
In particular, \citet{Jiang:15} simulated the radiation-dominated envelopes of massive stars in a plane parallel
geometry and found that in certain regions close to the surface of the star more shallow than a critical optical depth $\tau < \tau_\mathrm{crit}$, convection is unable to carry the super-Eddington flux at the iron opacity peak (which is convectively unstable and locally super-Eddington). When $\tau \gg \tau_\mathrm{crit}$, convection is efficient (in the radiative sense), with the averaged radiation advection flux consistent with the calculation of convection flux following MLT and sub-Eddington radiative acceleration. In contrast, radiation carries non-negligible flux for $\tau < \tau_\mathrm{crit}$, and a very significant fraction of the flux for $\tau \ll \tau_\mathrm{crit}$, despite the convective motion. Even while $\tau \ll \tau_\mathrm{crit}$, the envelope continues to behave in a turbulent and non-stationary manner, with shocks forming and driving large changes in density, and density inversions dissipating (due to convection) and reforming cyclically. The density inversion at the iron opacity peak is ultimately retained in the time-averaged density profile and the radiative acceleration is larger than the gravitational acceleration, creating an unbinding effect, though a mass loss or mass loss rate estimate were unable to be determined since the simulation neglected line driving and the global stellar geometry.

\citet{Jiang:18} simulated large wedges of radiation-dominated envelopes in spherical geometry, and also explored different regions of the H-R diagram compared to \citet{Jiang:15}. They found that a convective instability at the iron opacity peak can lead to the formation of a strong helium opacity peak. Upon formation, the helium opacity peak causes local radiative acceleration to exceed the gravitational acceleration by an order of magnitude, leading to most of the overlaying mass being unbound and thereby triggering an outburst up to an instantaneous mass loss rate of up to $0.05~M_\odot \ \mathrm{yr}^{-1}$ \citep[which is still $\sim 10-100$ times smaller than observed by e.g., ][]{Smith:2014}. Additionally, after the envelope has settled into a steady state, convection causes large envelope oscillations during which mass is lost at a rate of $\approx5 \times 10^{-6}~M_\odot \ \mathrm{yr}^{-1}$.  

Here, instead of enhancing energy transport artificially as in MLT++ \citep{paxton:13} and the implicit method \citep{Jermyn:2023}, we introduce a 1D model that converts local super-Eddington radiative luminosity into an average global outflow and explore the effect of this mass loss on stellar evolution.

\section{Methods} \label{sec:methods}

\subsection{MESA}

We used \texttt{MESA} version \texttt{r23.05.1}  \citep{paxton:11,paxton:13,paxton:15,paxton:18,paxton:19,Jermyn:2023} to evolve stars between the masses of $10-100~M_\odot$ (at every $1~M_\odot$ mass increment) at $1~Z_\odot=0.0142$ \citep[following][]{Houdek:2011, Asplund:2021} and $0.2~Z_\odot$ (SMC metallicity). We used the \citet{Grevesse:98} metallicity abundance ratios (re-scaled to $Z_\odot=0.0142$), and the \texttt{approx21.net} nuclear network \citep{Timmes:99}. Non-rotating models were allowed to evolve from the pre-main-sequence until the end of core helium burning (defined as central helium less than a mass fraction of $10^{-4}$). All models evolve through core helium ignition, but due to aforementioned numerical issues (see Section~\ref{sec:massive_stars}), not all models reach the end of core helium burning. 

For convection, we used the MLT treatment following \citet{Cox:68} with an $\alpha_\mathrm{MLT}$ of $1.73$ following \citet{Jones:2013, Li:2019} based on fits to the Sun \citep{Trampedach:2011}. We account for convective overshooting \citep{Herwig:00,paxton:11} with a combination of step and exponential overshooting for, respectively, the top of the hydrogen burning core (\texttt{overshoot\_f} $= 0.345$ and \texttt{overshoot\_f0} $= 0.01)$ and the top of all other convective cores (\texttt{overshoot\_f} $= 0.01$ and \texttt{overshoot\_f0} $= 0.005)$. This closely follows the setup used to produce Figure 18 in $\S7.2$ of \citet{Jermyn:2023}. Our inlists are provided at this repository:\dataset[DOI: 10.5281/zenodo.13306749]{https://doi.org/10.5281/zenodo.13306749}.

The \texttt{MESA} equation of state (EOS) combines OPAL \citep{Rogers2002}, SCVH
\citep{Saumon1995}, FreeEOS \citep{Irwin2004}, HELM \citep{Timmes2000},
PC \citep{Potekhin2010}, and Skye \citep{Jermyn:21} EOSes. Radiative opacities combines OPAL \citep{Iglesias1993,
Iglesias1996} and data from \citet{Ferguson2005} and \citet{Poutanen2017}.  Electron conduction opacities are from
\citet{Cassisi2007}. Nuclear reaction rates are from JINA REACLIB \citep{Cyburt2010}, NACRE \citep{Angulo1999} and \citet{Fuller1985, Oda1994,
Langanke2000}.  Screening is included via the prescription of \citet{Chugunov2007}.
Thermal neutrino loss rates are from \citet{Itoh1996}.

\subsection{Mass loss model}

In \texttt{MESA}, mass loss is implemented through wind mass-loss schemes. We employ \texttt{MESA}'s `\texttt{Dutch}' prescription for $T_\mathrm{eff}>4000~\mathrm{K}$, which combines the models of \cite{Vink:2001} and \cite{Nugis:2000} as described in \cite{Glebbeek:2009} and captures the mass loss rate due to line-driven stellar winds $\dot{M}_\mathrm{wind}$. For $T_\mathrm{eff} < 4000~\mathrm{K}$, we adopt the more recent empirical mass loss rates of RSGs from \citet{Decin:2024}. This replaces the \citet{deJager:88} or \citet{Nieuwenhuijzen:90} mass loss rates typically used in \texttt{MESA}'s `\texttt{Dutch}' prescription which are considered to overestimate mass loss rates (see Section~\ref{sec:intro}). The strength of the \texttt{Dutch} wind is modulated by the \texttt{Dutch\_scaling\_factor}, which is uncertain. We choose a \texttt{Dutch\_scaling\_factor} of $0.8$  \citep[following e.g.][]{Glebbeek:2009,Fields:21,Johnston:24, Pereira:24, Shiber:24} throughout our evolution.

To account for eruptive mass loss, we create a model based on the following energy argument (see also Section~\ref{sec:massive_stars}). Motivated by \citet{Jiang:15}, we define the transition between convectively-inefficient ($\tau<\tau_\mathrm{crit}$) and convection-dominated ($\tau>\tau_\mathrm{crit}$) energy transport at a critical optical depth $\tau_\mathrm{crit}$:
    \begin{equation}\label{eqn:taucrit}
        \tau_\mathrm{crit} \equiv \frac{c}{c_\mathrm{s}} \ \text{,}
    \end{equation}

\noindent where $c_\mathrm{s}$ is the total sound speed. This is similar to Equation~1 in \citet{Jiang:15}, except we adopt the total sound speed rather than the gas isothermal sound speed as a conservative estimate\footnote{$c_\mathrm{s} > c_\mathrm{gas, isothermal}$, so $\tau_\mathrm{crit}$ used here is smaller and therefore more conservative.} for the effective energy transport by dynamical processes. The critical optical depth $\tau_\mathrm{crit}$ can be found by equating the photon diffusion timescale and the acoustic timescale for a radial distance $\Delta{r}$ in the star, for which \begin{align}
\text{acoustic timescale} & = \frac{\Delta{r}}{c_s}  \\
\text{photon diffusion timescale} & = \frac{\Delta{r}}{c/\tau}
\end{align} where $c_s$, $\Delta{r}$, and $\tau$ are local values of, respectively, total sound speed, radial distance, and optical depth. Thus, $\tau_\mathrm{crit}$ is a local value determined at every radius. In the stellar envelope, $\tau = \tau_\mathrm{crit}$ is only true at one point since both $c_s$ and $\tau$ are monotonic functions, and therefore $\tau = \tau_\mathrm{crit}$ defines a unique location above which $\tau < \tau_\mathrm{crit}$ everywhere in the outer envelope.

We also follow \citet{Jiang:15} and \citet{Jiang:18}'s interpretation of the balance between radiative acceleration and gravitational acceleration in driving outbursts of mass loss. Specifically, in the $\tau<\tau_\mathrm{crit}$ region, convection transports the convective luminosity $L_\mathrm{conv}$, and radiative diffusion transports the radiative luminosity $L_\mathrm{rad}$, but only up to the Eddington limit $L_\mathrm{Edd}$, defined as:
\begin{equation}\label{eqn:Eddington}
    L_\mathrm{Edd} = \frac{4\pi G c M_\mathrm{enc}}{\kappa} \ \text{,}
\end{equation}

\noindent with $M_\mathrm{enc}$ and $\kappa(r)$ being, respectively, the enclosed mass and local opacity.

\begin{figure}[ht!]
\centering
\includegraphics[width=1.0\columnwidth]{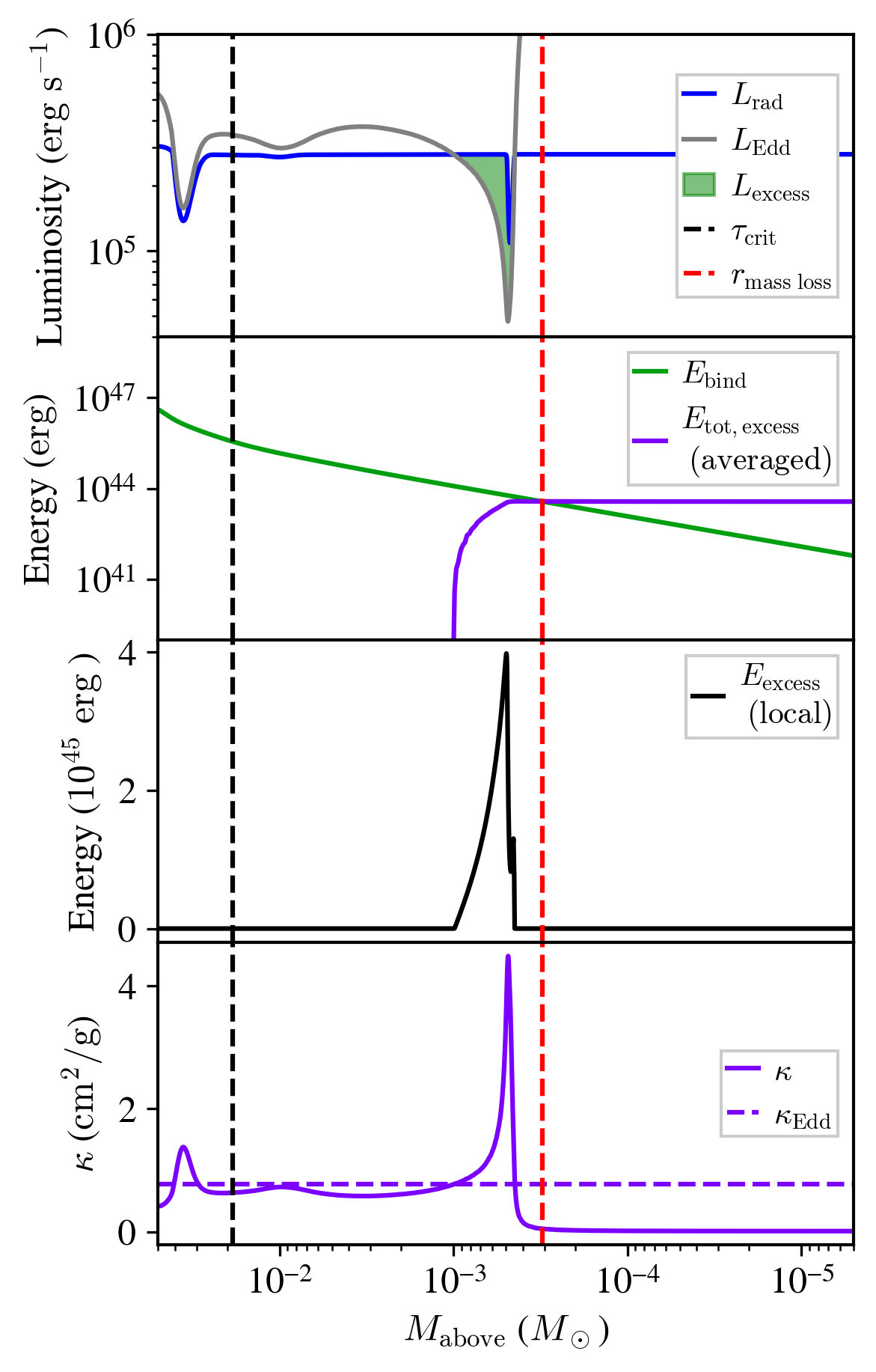}
\caption{Profiles of luminosity, energy, and opacity as a function of mass above each radius for a $M_\mathrm{ZAMS}=30~M_\odot$, $Z=1~Z_\odot$ star. The model chosen corresponds to time when $\log T_\mathrm{eff}/\mathrm{K} = 3.76$, $\log L/L_\odot = 5.45$ on the H-R diagram just after the start of core helium burning (center helium mass fraction of $0.98$). For all panels, the horizontal axis is defined as $M_\mathrm{above} = M - m(r)$. The dashed black 
and red lines respectively show $\tau_\mathrm{crit}$ and $r_\mathrm{mass\ loss}$ as defined in Equations~\ref{eqn:taucrit} and \ref{eqn:rmassloss}. The x-intercept of the red dashed line shows the total mass lost (see Equation~\ref{eqn:Mtot}). Top panel: local radiative, Eddington (Equation~\ref{eqn:Eddington}), and excess luminosities. Second panel: total excess energy (Equation~\ref{eqn:Etotexcess}) and binding energy (Equation~\ref{eqn:Ebind}). Third panel: local excess energy (Equation~\ref{eqn:Eexcess}). Last panel: local opacity, showing the opacity peak at around $M_\mathrm{above} \approx 4 \times 10^{-4}~M_\odot$ due to both hydrogen and helium ionization transitions. }\label{fig:mabove}
\end{figure}

\begin{figure*}[ht!]
    \centering
\includegraphics[width=1.0\textwidth]{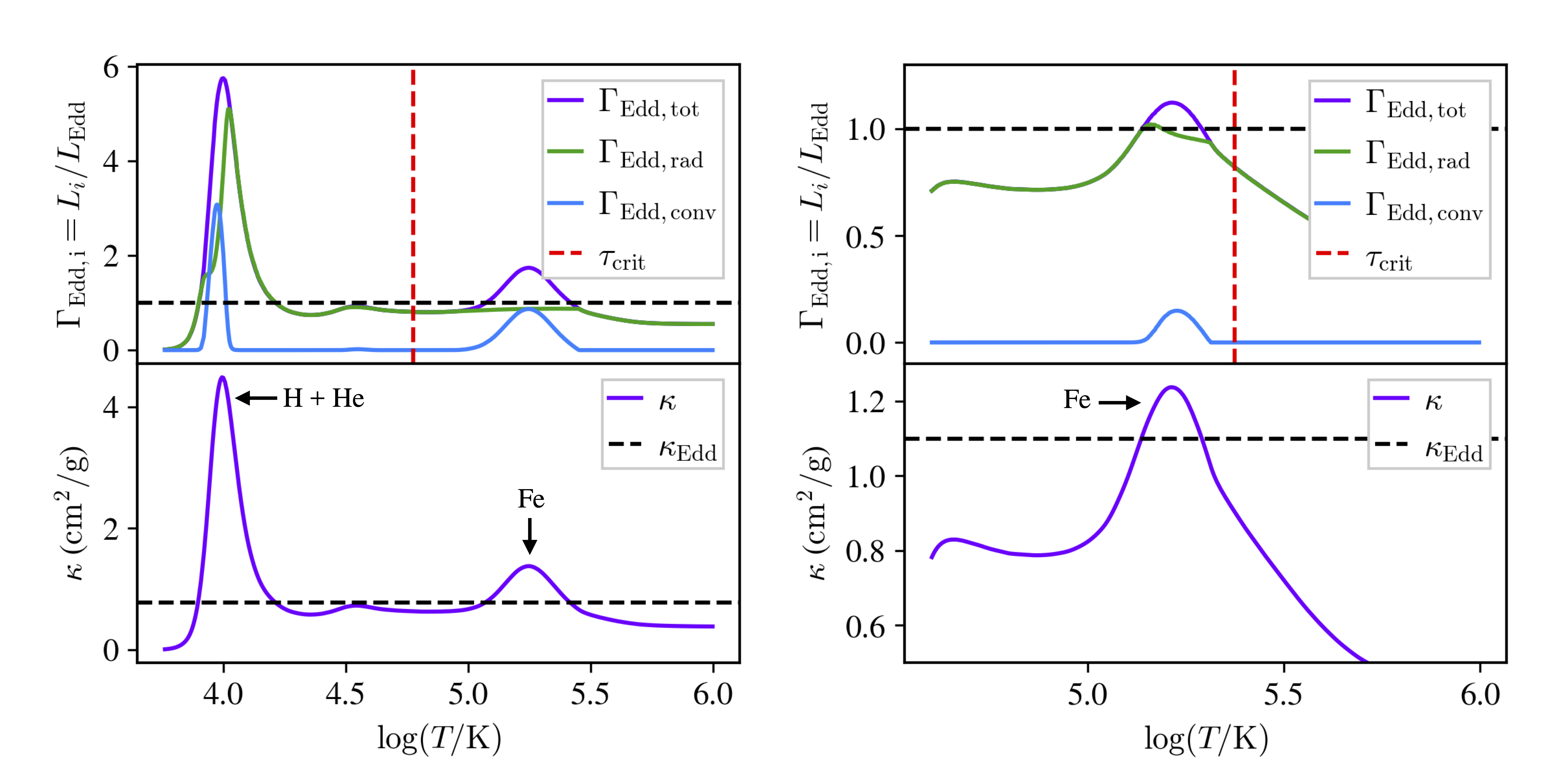}
    \vspace{-20pt}\caption{Stellar temperature profiles showing locally super-Eddington radiative luminosity due to opacity peaks. Left panels: a $M_\mathrm{ZAMS}=30~M_\odot$, $Z=1~Z_\odot$ model with a opacity peak due to both hydrogen and helium ionization transitions. Left panels correspond to time when $\log T_\mathrm{eff}/\mathrm{K} = 3.76$, $\log L/L_\odot = 5.45$ on the H-R diagram. Right panels: a $M_\mathrm{ZAMS}=60~M_\odot$, $Z=1~Z_\odot$ model with an iron opacity peak. Right panels correspond to time when $\log T_\mathrm{eff}/\mathrm{K} = 4.60$, $\log L/L_\odot = 5.80$ on the H-R diagram. Top panels: Eddington ratios; ratios of radiative (purple) and convective (green) luminosity against Eddington luminosity. Here, radiative luminosity or convective luminosity refer to, respectively, the luminosity carried by radiative diffusion or convection. Unity is shown in a dashed black line, and the critical optical depth is indicated by a dashed red line (see text in Section~\ref{sec:methods}.) Bottom panels: local opacity in each zone throughout the star (purple) and Eddington opacity (dashed black) $4\pi G c M/L$. Only locally super-Eddington radiative luminosity (in green) is considered in our eruptive mass loss model.}\label{fig:opacity_peaks}\vspace{0.3pt}
\end{figure*}

\begin{figure}[!htb]
    \centering
    \gridline{
          \fig{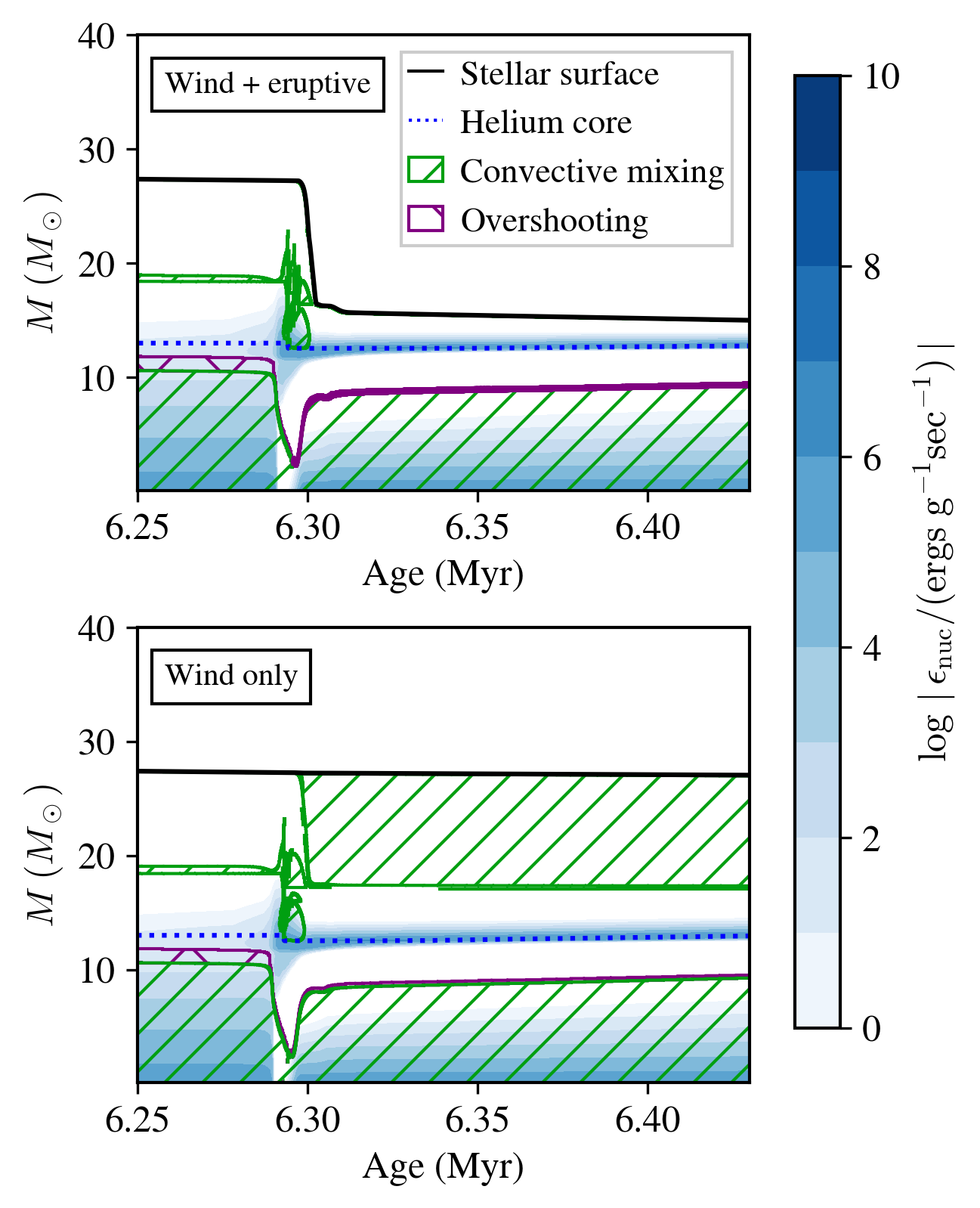}{\linewidth}{}
          }
        \vspace{-25pt}
        \caption{Kippenhahn diagram for $M_\mathrm{ZAMS}=30~M_\odot$, $Z=1~Z_\odot$ stars. The age x-axis focuses on the time of core He burning. This is the same model shown in Figure~\ref{fig:mabove}. Top panel: modeled with stellar wind and eruptive mass loss. Bottom panel: modeled only with stellar wind. Blue contours show the rate of nuclear energy generation, with colorbar on the right. } \label{fig:kipp}
\end{figure}

For any locally super-Eddington radiative luminosity (i.e., for any $L_\mathrm{rad}$ exceeding $L_\mathrm{Edd}$), we regard the excess radiative luminosity above $L_\mathrm{Edd}$ as available to do work to unbind the envelope and contribute to eruptive mass loss. More formally, since photons are able to couple with matter when $\tau \gtrsim 2/3$, the $2/3 \lesssim \tau<\tau_\mathrm{crit}$ region is of interest if the Eddington limit is locally exceed by the radiative luminosity. In this case, the photons in the radiatively super-Eddington regions in $2/3 \lesssim \tau<\tau_\mathrm{crit}$ can energetically couple with matter and drive mass loss. Therefore we consider any excess radiative luminosity above Eddington ($L_\mathrm{excess} = L_\mathrm{rad}-L_\mathrm{Edd} > 0$) in regions of the stellar model with $\tau < \tau_\mathrm{crit}$ as available to drive mass loss. The ratio of the radiative, convective, and total luminosity to the Eddington luminosity can be expressed as $\Gamma_\mathrm{Edd, rad} = L_\mathrm{rad}/L_\mathrm{Edd}$, $\Gamma_\mathrm{Edd, conv} = L_\mathrm{conv}/L_\mathrm{Edd}$, and $\Gamma_\mathrm{Edd, tot} = L_\mathrm{tot}/L_\mathrm{Edd}$ respectively.

We choose to adopt the local dynamical time $t_\mathrm{dyn}$, since eruptive mass loss must be dynamical, to define a local excess energy $E_\mathrm{excess}$ as 
    \begin{equation}\label{eqn:Eexcess}
        E_\mathrm{excess} = E_\mathrm{excess}(r) = L_\mathrm{excess} t_\mathrm{dyn} \ \text{,}
    \end{equation}
\noindent where $t_\mathrm{dyn}$ is
    \begin{equation}
        t_\mathrm{dyn} = t_\mathrm{dyn}(r) = \sqrt{\frac{r^3}{G m(r)}} \ \text{,}
    \end{equation}
\noindent and $r$ and $m(r)$ are, respectively, the radius and mass coordinates. $E_\mathrm{excess}$ is a local value at each radius $r$ of the star.

For $\tau < \tau_\mathrm{crit}$, we find the total excess energy at every radius $r$ from
    \begin{equation}\label{eqn:Etotexcess}
        E_\mathrm{tot, excess} = E_\mathrm{tot, excess}(r) = \frac{\int_{m(r(\tau_\mathrm{crit}))}^{m(r)} E_\text{excess} \ dm }{\int_{m(r(\tau_\mathrm{crit}))}^{m(r)}{dm}}\ \text{.}
    \end{equation}

This total excess energy is a cumulative integral from $\tau_\mathrm{crit}$ outwards, and represents a mass-average of the excess energy in the super-Eddington region. $E_\mathrm{tot, excess}$ has a value at every radius $r$ up to the surface and is only defined for $\tau < \tau_\mathrm{crit}$, since this is the only region we consider as possibly able to drive mass loss (see Section~\ref{sec:massive_stars}). We note that the meshing in \texttt{MESA} is very close to uniform in the region of interest ($\tau < \tau_\mathrm{crit}$), which renders the weighting effectively irrelevant.

We compare this total excess energy with the gravitational binding energy of the overlaying material, which for the mass external to radius $r$ is expressed as
    \begin{equation}\label{eqn:Ebind}
        E_\mathrm{bind} = E_\mathrm{bind}(r) = -\int_{m(r)}^{M}{\frac{Gm}{r} dm} \ \text{,}
    \end{equation}

\noindent for which $G$ is the gravitational constant, $m$ and $r$ are the local mass and radius, $dm$ is the local mass increment in question (at radius $r$), and $M$ is the mass at the surface (total mass). We conservatively choose not to include other sources of energy such as thermal energy, recombination energy, and kinetic energy, which are all positive energies that decrease $E_\mathrm{bind}$. Thus, we expect our estimated eruptive mass loss rates to be conservative. 

For every radius $r$ where $\tau < \tau_\mathrm{crit}$, we define a local energy difference between the binding energy and the total excess energy as
    \begin{equation}\label{eqn:Ediff}
        E_\mathrm{diff} = E_\mathrm{diff}(r) = E_\mathrm{tot, excess} - |E_\mathrm{bind}| \ \text{.}
    \end{equation}
We find the inner-most radius $r_\mathrm{mass \ loss}$ greater than the radius of $\tau_\mathrm{crit}$ above which $E_\mathrm{diff} > 0$, and then find the difference between the surface mass $M$ and the mass at $r_\mathrm{mass\ loss}$ to give $\Delta{M}$. This can be expressed as
    \begin{equation}\label{eqn:Mtot}
        \Delta{M} = M - m(r_\mathrm{mass \ loss}) \ \text{,} 
    \end{equation}
\noindent where
\begin{equation}\label{eqn:rmassloss}
r_\mathrm{mass\ loss}  \equiv  \min(r) > r(\tau_\mathrm{crit}) \quad \text{for which} \quad E_\mathrm{diff} > 0  \ \text{.}
\end{equation}

We emphasize that, at each timestep, $E_\mathrm{bind}$, $E_\mathrm{tot, excess}$, and $E_\mathrm{diff}$ are functions of radius $r$. We show how local luminosities (radiative, Eddington, and excess), local excess energy, and total excess energy (a cumulative integral) are accounted for when determining mass loss for our model in Figure~\ref{fig:mabove}, which shows a representative model of a $M_\mathrm{ZAMS}=30~M_\odot$ and $Z=1~Z_\odot$ star with mass loss due to a hydrogen/helium opacity peak (also shown in the left panel of Figure~\ref{fig:opacity_peaks}). Figure~\ref{fig:opacity_peaks} shows examples of how the opacity peaks (lower panels) can cause the star's radiative luminosity to locally exceed the Eddington limit (green line in top panels).

Finally, we define the mass loss rate as

    \begin{equation}
        \dot{M}_\mathrm{erupt} = \xi \frac{\Delta{M}}{t_\mathrm{dyn, local}} \ \text{,}
    \end{equation}

\noindent where $\xi$ is a free parameter we introduce to tune our model. $t_\mathrm{dyn, local}$ is the local dynamical time at $r_\mathrm{mass\ loss}$; this is the local dynamical time at the smallest radius in the star for which both $E_\mathrm{tot, excess} >  |E_\mathrm{bind}|$ (where the purple line exceeds the green line in the second panel of Figure~\ref{fig:mabove}) and $\tau < \tau_\mathrm{crit}$ (to the right of the dashed black line in Figure~\ref{fig:mabove}). We explore two values of $\xi$ in our modelling; $\xi=0.1$ and $\xi=1.0$.

For our total mass loss model, we combine the wind mass-loss model with our novel eruptive mass loss model:
\begin{equation}
    \dot{M} = \dot{M}_\mathrm{erupt} + \dot{M}_\mathrm{wind} \ \text{.}
\end{equation}

This mass loss model is implemented as a custom hook in \texttt{run\_star\_extras.f90} and is calculated at each timestep of a star's evolution. We note that although no explicit metallicity dependence is included in our eruptive mass loss model, the mass loss rate is expected to be indirectly impacted by changes in stellar structure due to metallicity \citep[e.g.,][]{Xin22} such as the effective temperature as well as the depth and strength of the helium and iron opacity peaks.

Using this model, we find that our eruptive mass loss model behaves as expected; mass loss occurs due to locally super-Eddington radiative luminosity at the hydrogen/helium and iron opacity peaks. As shown in Figure~\ref{fig:opacity_peaks}, these opacity peaks occur in the $\tau < \tau_\mathrm{crit}$ region (left of the red dashed line). In particular, the radiative luminosity (marked in green lines and denoted by $\Gamma_\mathrm{Edd, rad}$) can be locally super-Eddington (above the black dashed line) and correspond to the locations of opacity peaks. This locally super-Eddington radiative luminosity can drive eruptive mass loss and, when $\dot{M}_\mathrm{erupt}$ is sufficiently large, can alter the structure of the star by preventing the formation of an outer convective envelope. 

\section{Results} \label{sec:results}

\begin{figure}[t!]
        \includegraphics[width=0.9\columnwidth]{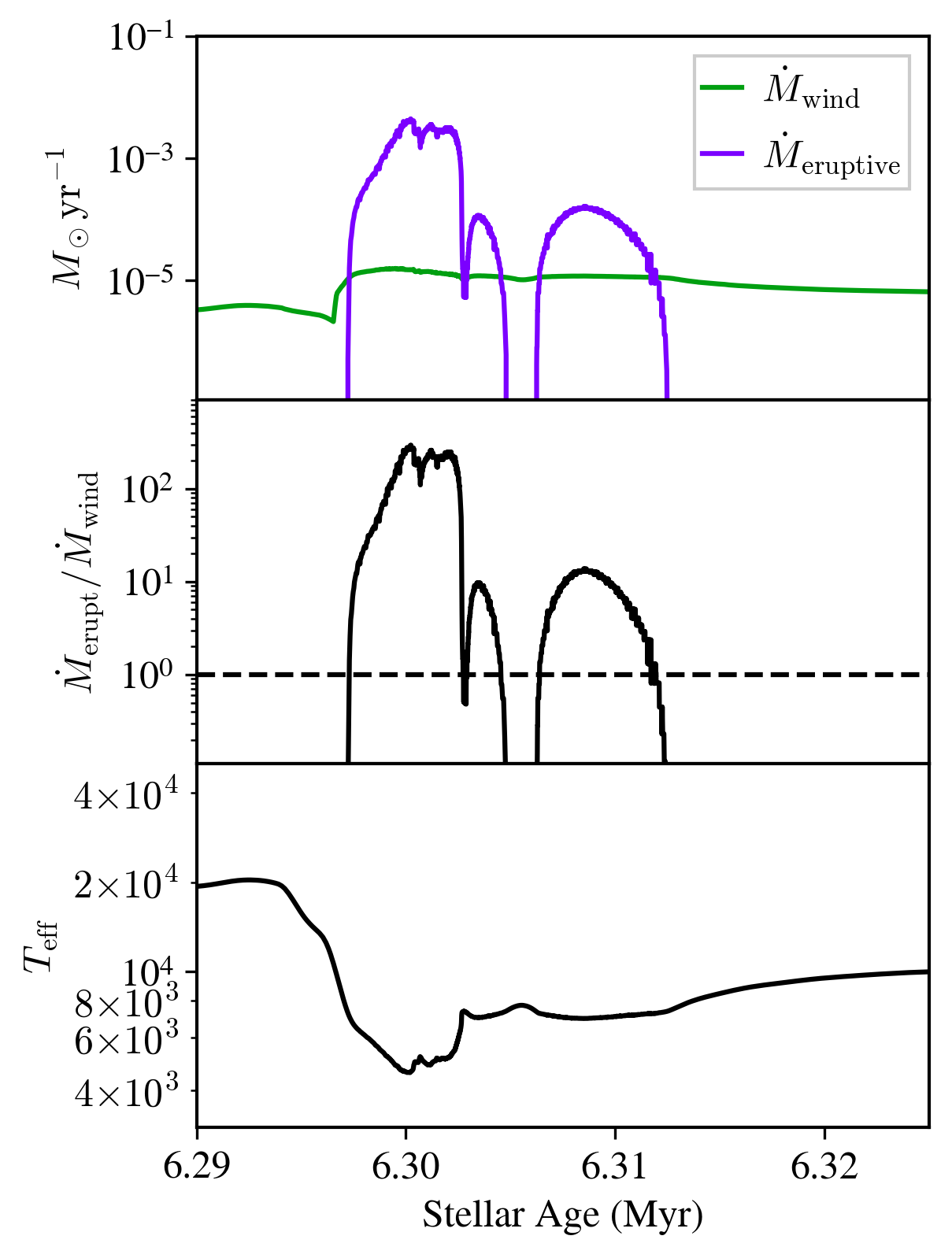}
        \caption{Mass loss rate, mass loss ratio, and effective temperature over time of a $M_\mathrm{ZAMS}=30~M_\odot$ star with $\xi=1.0$ and $Z=1~Z_\odot$. The middle panel shows the ratio between the mass loss rate due to eruptive mass loss and the stellar wind mass loss rate. The dashed black 
line marks when the ratio is unity.}\label{fig:masslosstime}
\end{figure}

    \begin{figure*}[ht!]
    \gridline{\fig{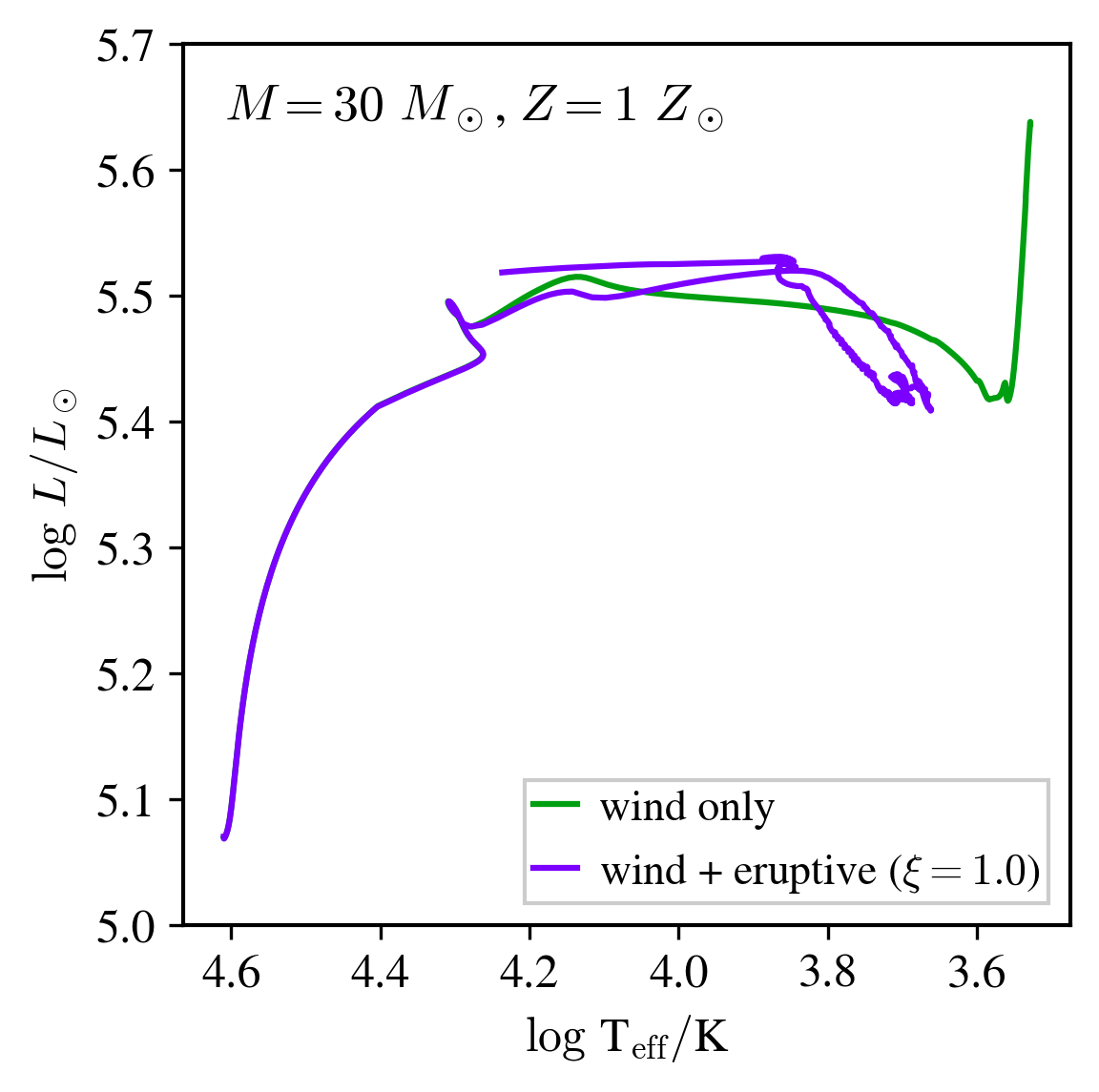}{0.48\textwidth}{}
          \fig{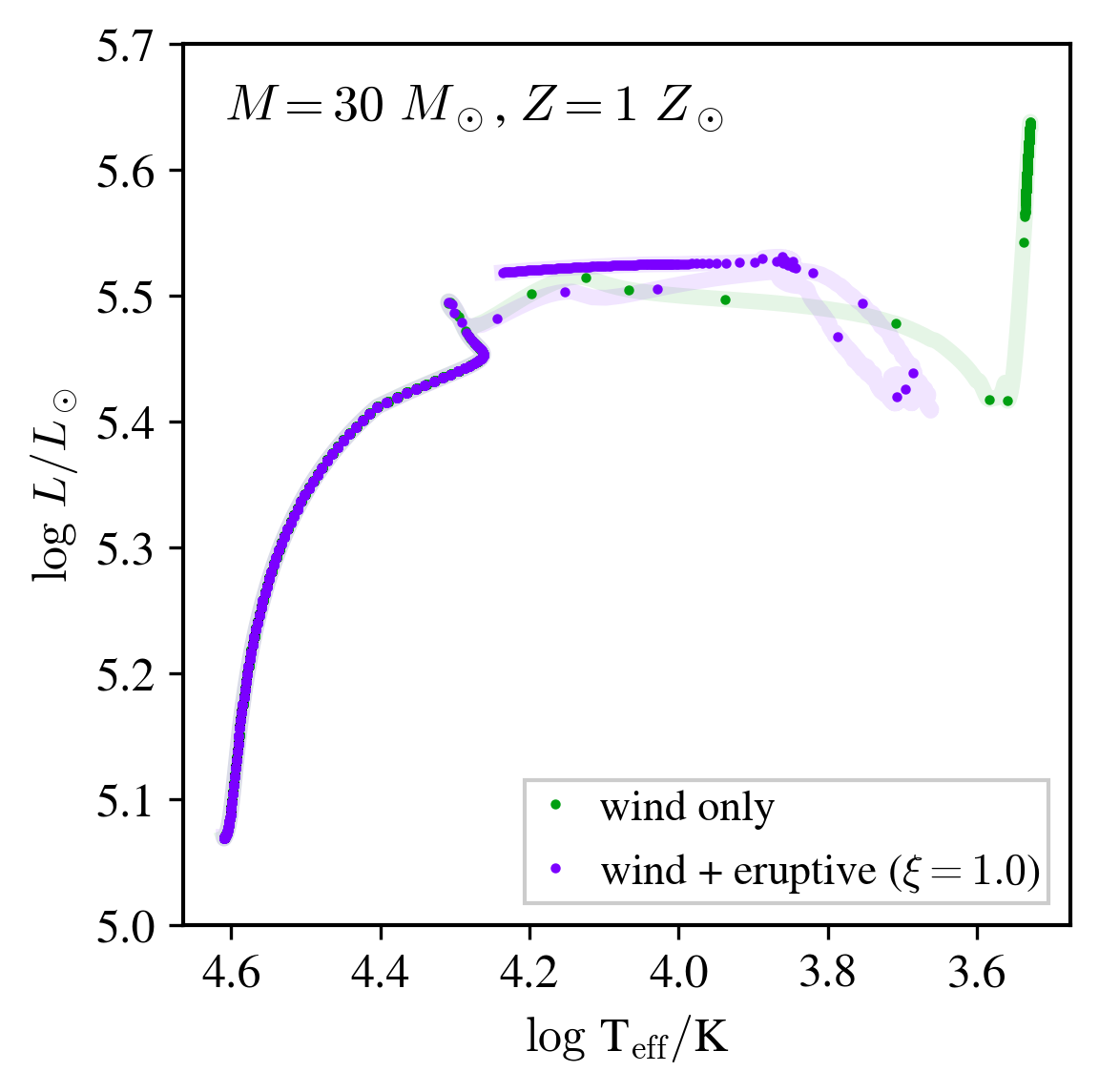}{0.48\textwidth}{}
          }
        \vspace{-20pt}
        \caption{Evolutionary tracks of $M_\mathrm{ZAMS}=30~M_\odot$, $Z=1~Z_\odot$ stars. Left panel: H-R tracks for wind only (green) and eruptive mass loss $+$ wind (purple) models. From the differing green and purple tracks, eruptive mass loss prevents evolution to RSG phase. Right panel: evolutionary tracks sampled to 1 model every $1000~\mathrm{years}$ to show where stars are expected to be observed.}\label{fig:tracks}
\end{figure*}

In this section, we show our results for eruptive mass loss. Figure~\ref{fig:kipp} presents Kippenhahn diagrams (mass profile over time) of stars modelled with and without eruptive mass loss in addition to stellar winds. As shown in the green hatches in the bottom panel for the star modelled without eruptive mass loss, a convective H-rich envelope is present above the helium-burning region. However, for the model with eruptive mass loss, this convective layer never forms due to eruptive mass loss at $6.3~\mathrm{Myr}$ \citep[in agreement with][]{Schootemeijer:24}. 

Figure~\ref{fig:masslosstime} more clearly shows the timing of this eruptive mass loss by presenting the mass loss rates due to stellar winds and eruptive mass loss over time, as well as the relative importance of $\dot{M}_\mathrm{erupt}$ compared to $\dot{M}_\mathrm{wind}$. We see that the mass loss at around $6.3~\mathrm{Myr}$ is indeed primarily due to eruptive mass loss, with $\dot{M}_\mathrm{erupt}$ exceeding $\dot{M}_\mathrm{wind}$ by two orders of magnitude.

As a result of this eruptive mass loss, the convective layer is not formed and the star no longer evolves cool enough to become a RSG. This can be seen in Figure~\ref{fig:tracks}, which shows the evolutionary tracks of the two stars featured in Figure~\ref{fig:kipp} on a H-R diagram. For the model with eruptive mass loss (in purple), the star remains hotter throughout its evolution and does not evolve to become a RSG. This is contrasted with the model with only stellar winds (in green), which evolves cooler to become a RSG. The left panel of Figure~\ref{fig:tracks} compares models run with (purple) and without (green) eruptive mass loss in addition to stellar winds. The right panel of Figure~\ref{fig:tracks} samples the evolutionary track every $1000~\mathrm{years}$, which allows better correspondence to where we expect stars to be observed. Based on this, we can see that the star modelled with eruptive mass loss is likely to be observed as a blue supergiant (BSG) whereas the star modelled with only stellar winds is likely to be observed as a RSG. Additionally, these sampled tracks capture how the star evolves across the Hertzsprung gap very rapidly in $\sim 10,000~\mathrm{years}$. We note that our results are not sensitive to the choice of RSG wind model since eruptive mass loss prevents evolution to the low temperatures applicable to RSG wind models.

\begin{figure*}
    \centering
    \gridline{\fig{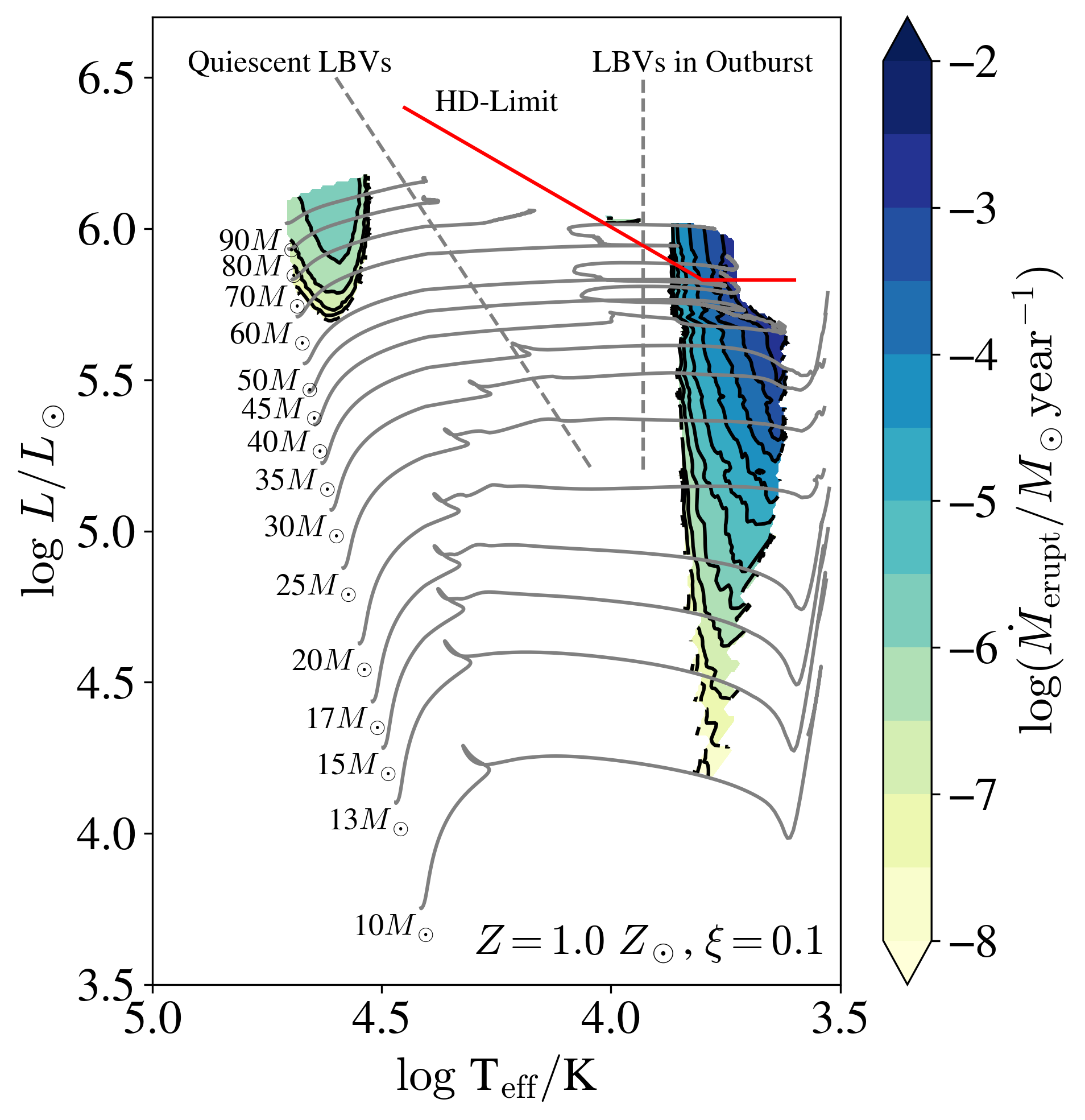}{0.5\textwidth}{}
          \fig{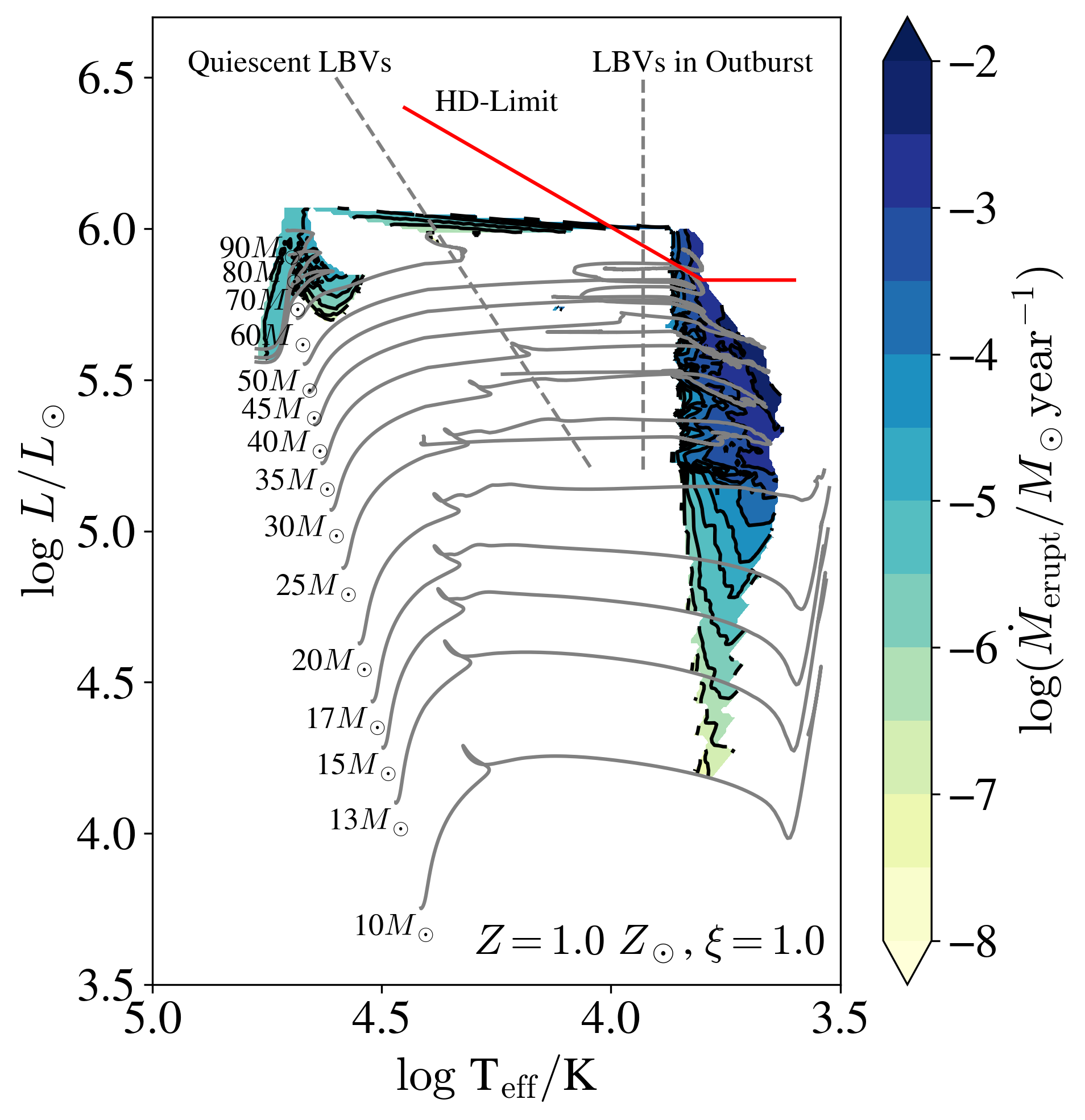}{0.5\textwidth}{}
          }
    \gridline{\fig{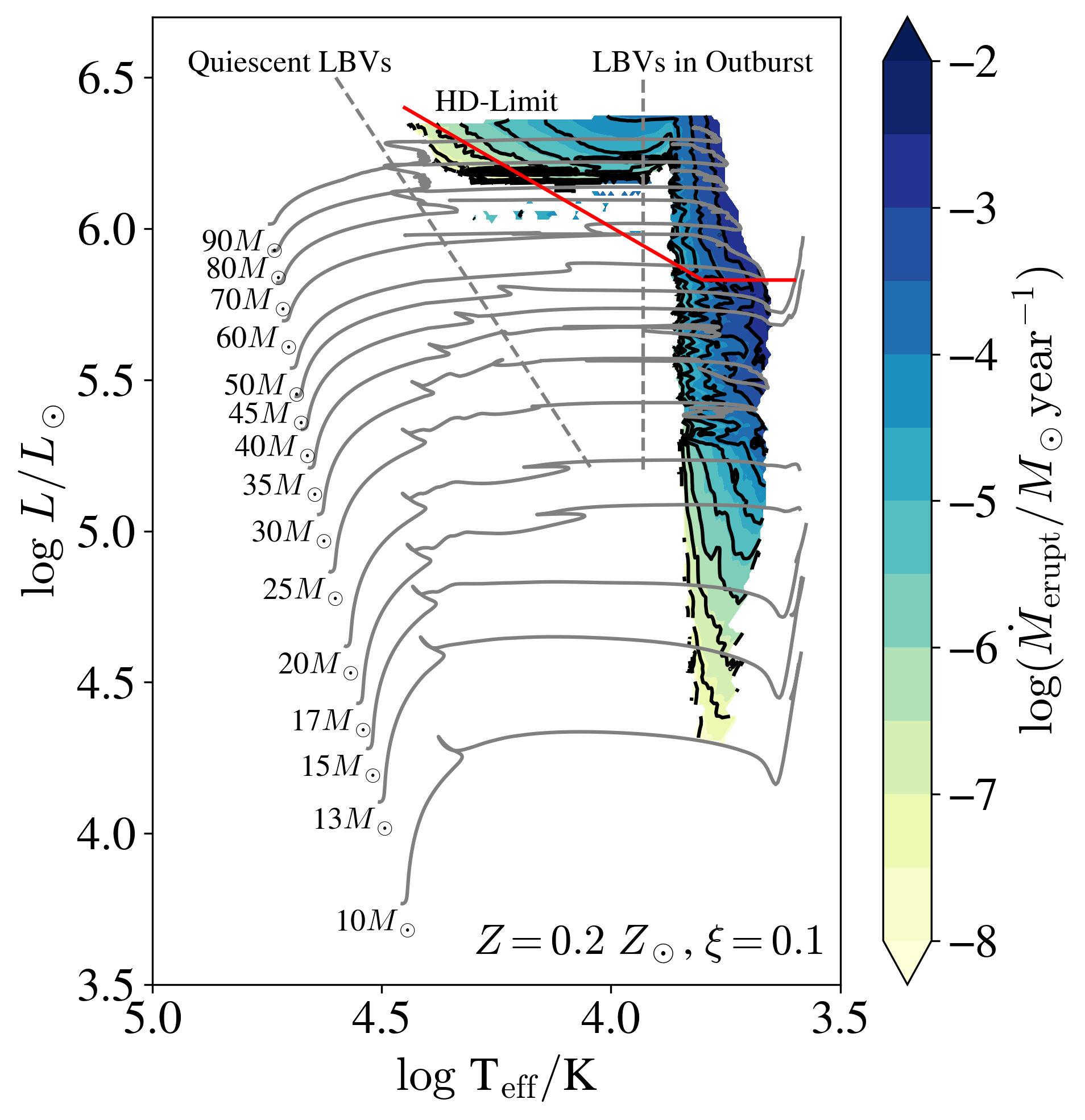}{0.5\textwidth}{}
          \fig{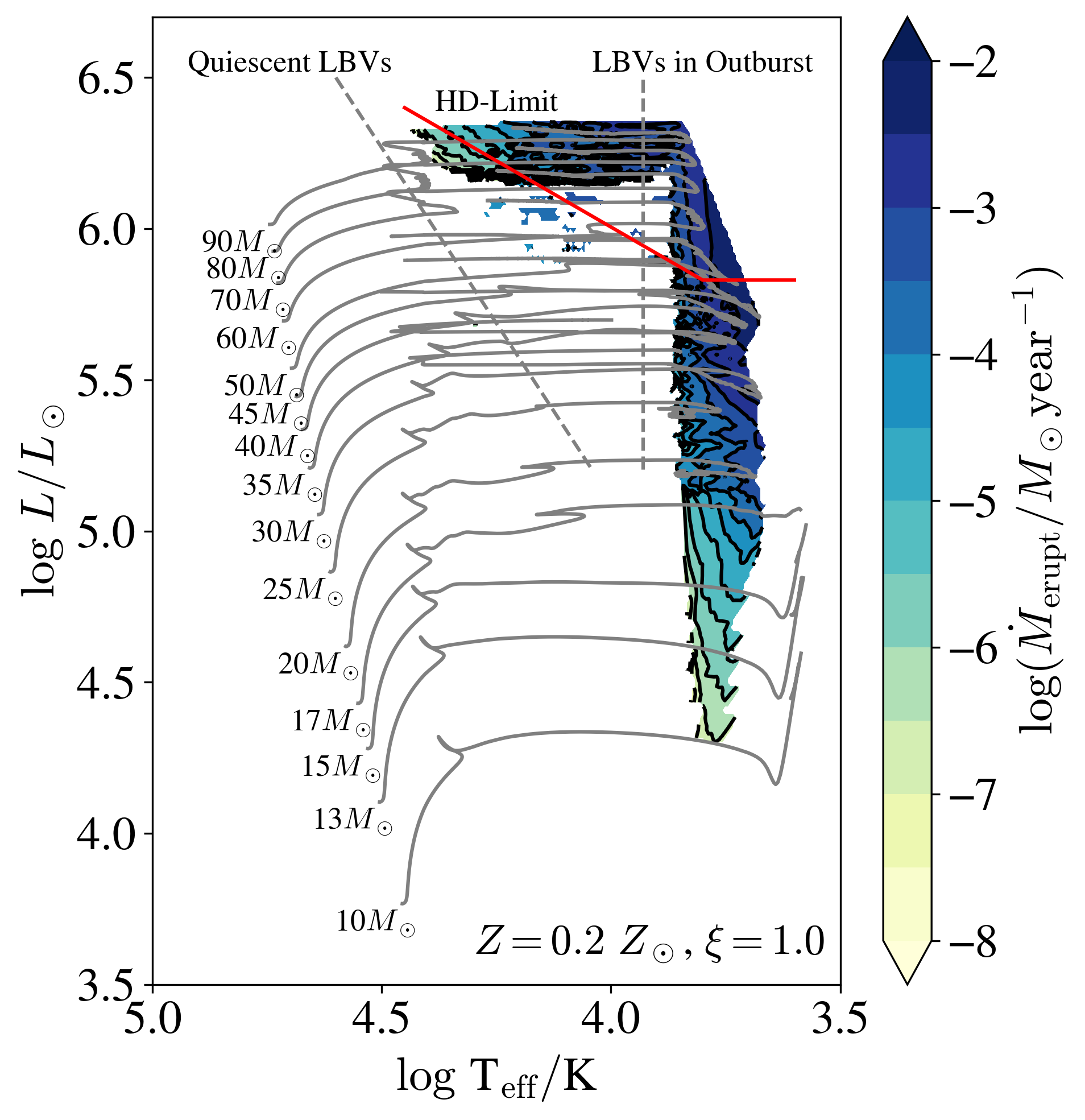}{0.5\textwidth}{}
          }
    \caption{Eruptive mass loss rates across the H-R diagram. Left column: $\xi=0.1$. Right column: $\xi=1.0$. Top row: $Z=1~Z_\odot$. Bottom row: $Z=Z_\mathrm{SMC}=0.2~Z_\odot$. Grey lines show the evolutionary tracks of stars from the beginning of main sequence until the end of the model run. Models were run at every $1~M_\odot$ increments but only selected models are shown here for clarity. Dashed grey lines show two empirical tracks of Quiescent LBVs and LBVs in Outburst for reference. Solid red lines show the Humphreys-Davidson limit for reference. Colored contours show the rate of eruptive mass loss experienced by the star, with legend on the right.}\label{fig:res}
\end{figure*}

We present our full mass loss results for our $10-100~M_\odot$ stars at $Z=1~Z_\odot$ in the top panels of Figure~\ref{fig:res}, which shows the evolution tracks of stars that span the full mass range on an H-R diagram. The mass loss rate due to eruptive mass loss is colored in yellow, green, and blue contours. We take particular note that significant mass loss (green and blue contours) occurs in a vertical band between $3.5< \log(T_\mathrm{eff}/\mathrm{K}) < 4.0$ as the stars evolve past the main sequence towards the red supergiant phase. Most of these models experiencing eruptive mass loss retain a surface hydrogen mass fraction between $0.25-0.5$, with surface helium and nitrogen mass fractions between $0.5-0.75$ and $1-1.8\times10^{-3}$ respectively. The mass loss in this band is caused by the opacity peak in the stellar envelope due to both hydrogen and helium ionization transitions, as shown in Figure~\ref{fig:opacity_peaks} where the peaks in Eddington ratios coincide with the hydrogen/helium opacity peak at $\sim 10^4$~K \citep[see discussion in][]{Jiang:18}. This result agrees with the 3D models in \citet{Jiang:18} where helium opacity peaks in the envelope of massive stars were found to drive mass loss.

\begin{figure*}
    \centering
    \gridline{\fig{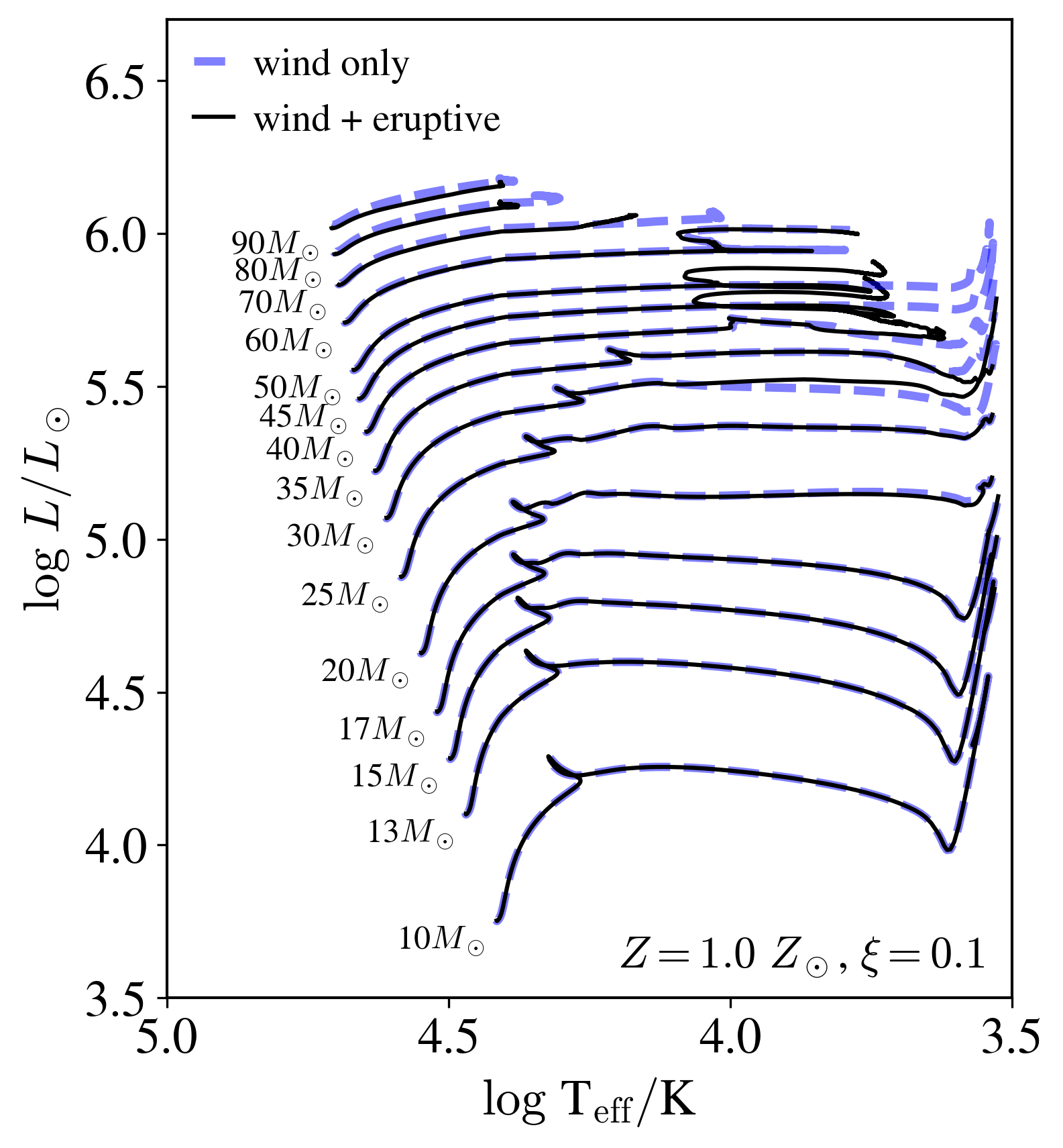}{0.48\textwidth}{}
          \fig{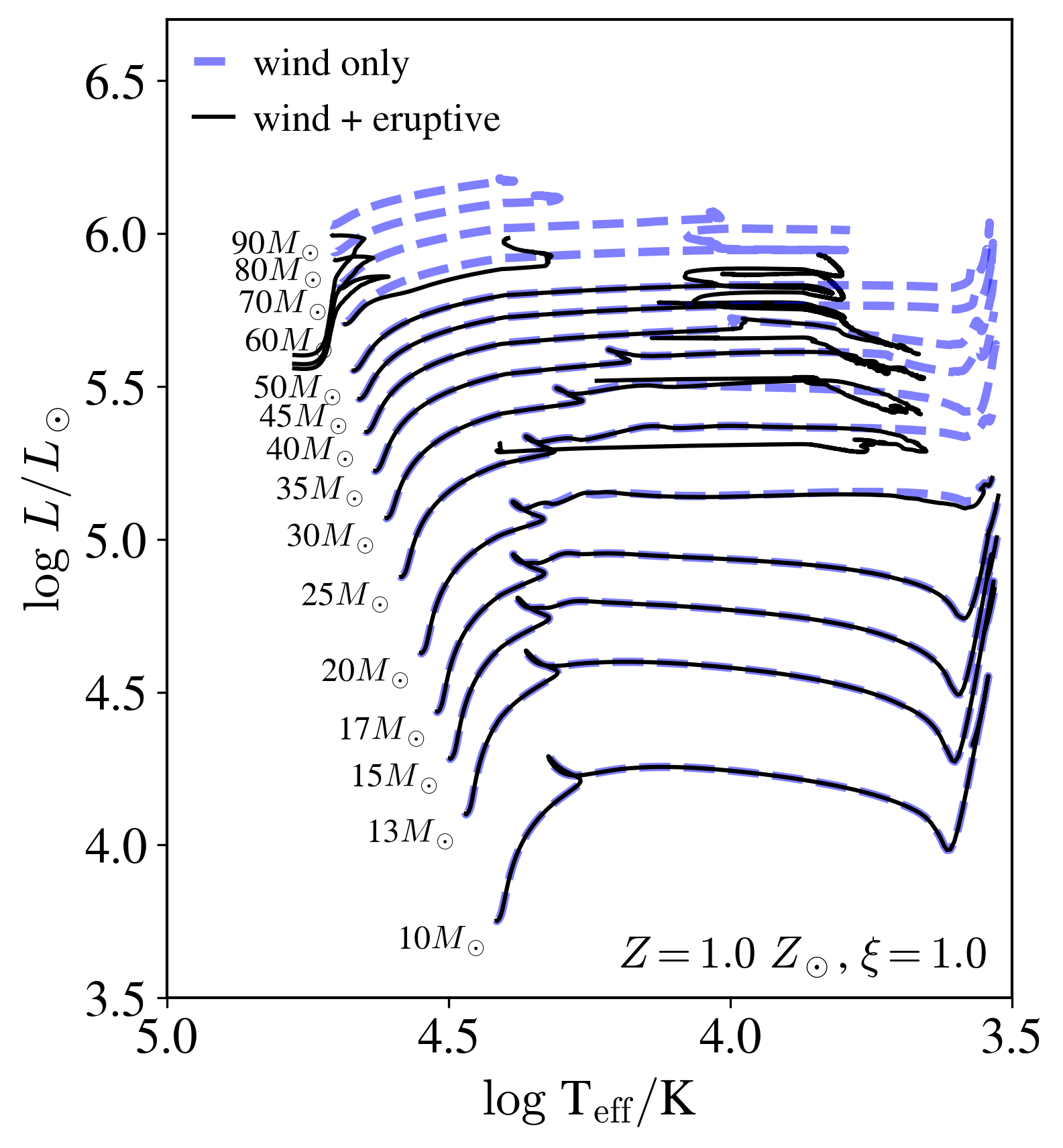}{0.48\textwidth}{}
          }
    \gridline{\fig{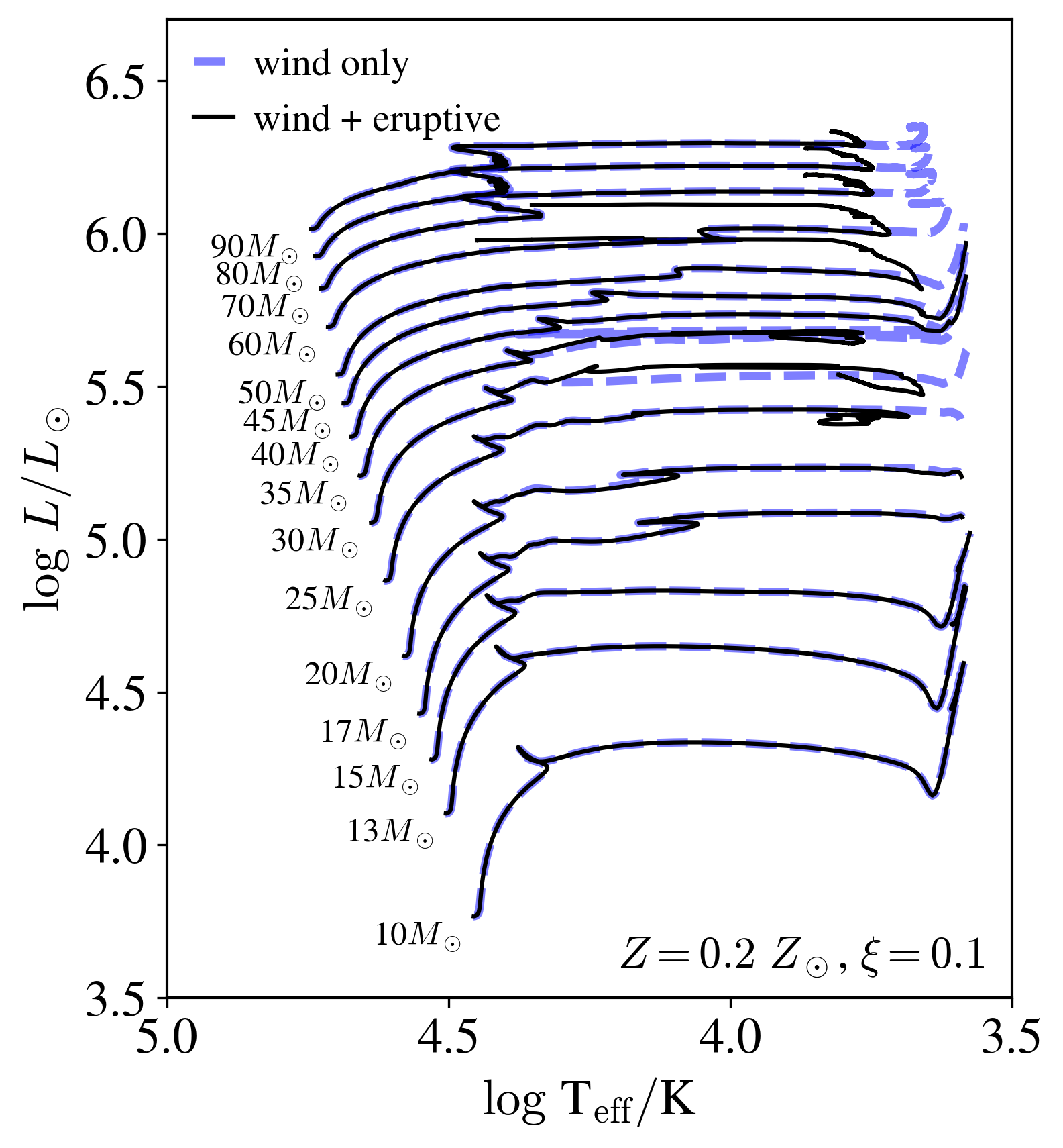}{0.48\textwidth}{}
          \fig{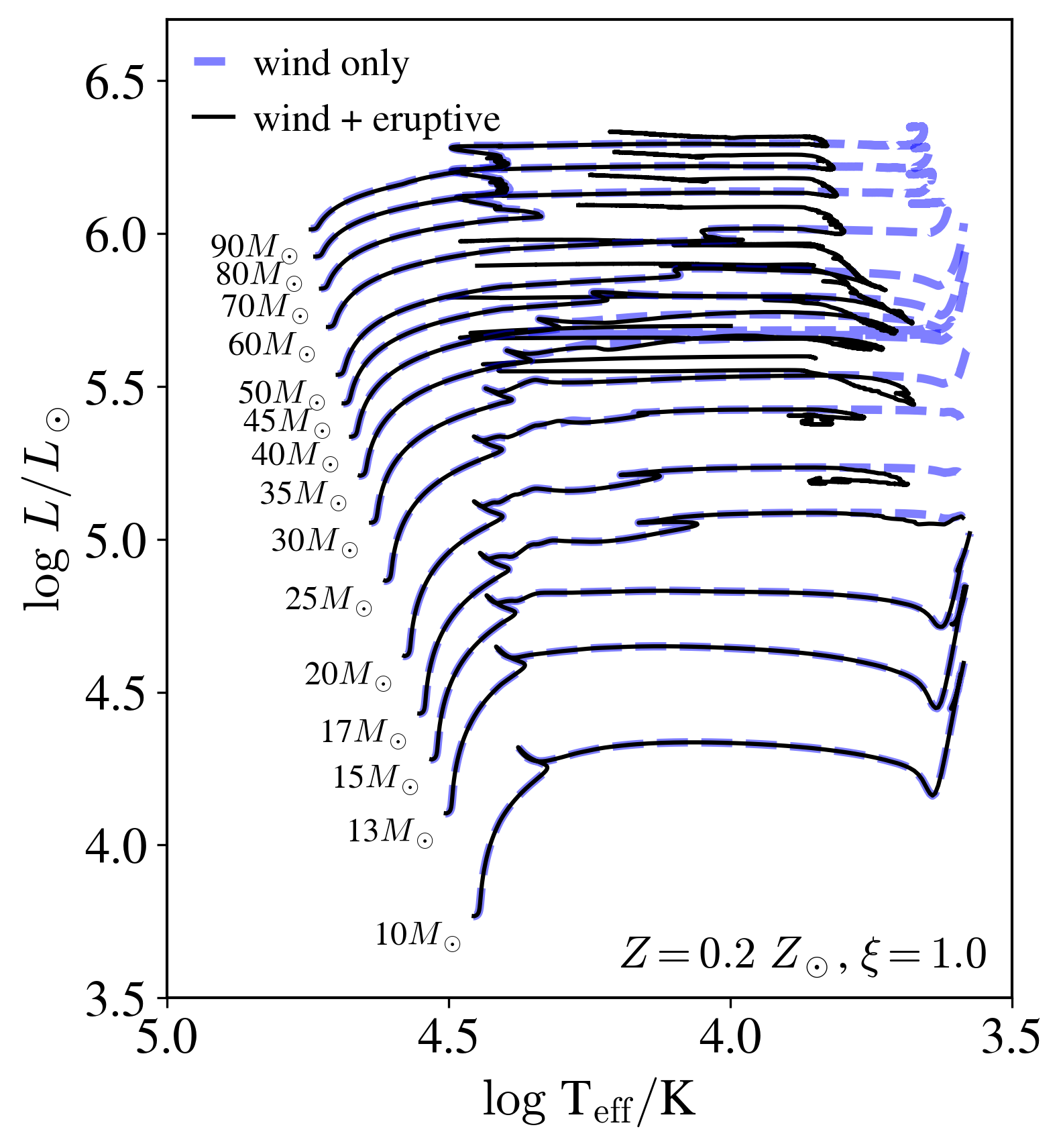}{0.48\textwidth}{}
          }
    \caption{Evolutionary tracks of models with and without eruptive mass loss. Left column: $\xi=0.1$. Right column: $\xi=1.0$. First row: $Z=1~Z_\odot$. Second row: $Z=Z_\mathrm{SMC}=0.2~Z_\odot$. All axes and labels follow Figure~\ref{fig:res}.}\label{fig:trackres}
\end{figure*}

Comparing the left and right panels in Figure~\ref{fig:res}, we see that while all the stars evolved with $\xi = 0.1$ (left column) evolve to become RSGs, stars with mass $\gtrsim20~M_\odot$ evolved with $\xi = 1.0$ (right column) do not become RSGs. This indicates that the presence of eruptive mass loss significantly alters the evolution of stars with mass $>20~M_\odot$, suggesting that eruptive mass loss may be a physical mechanism behind the lack of RSGs above $20-30~M_\odot$ as proposed by \citet{Massey:2000, Heger:2003, Levesque:2009,Ekstrom:12} and the excess of BSGs \citep{Bellinger:23}. We show the change in the evolutionary tracks of our stars modeled with eruptive mass loss compared to those modeled only with stellar wind in Figure~\ref{fig:trackres}. From the top right panel of Figure~\ref{fig:trackres}, we confirm that for $M_\mathrm{ZAMS}\gtrsim20~M_\odot$, models evolved with eruptive mass loss at $\xi=1.0$ do not evolve to temperatures cooler than $\log(T_\mathrm{eff}/\mathrm{K})\approx 3.65$ and therefore do not become RSGs.

\begin{figure*}
    \centering
    \gridline{\fig{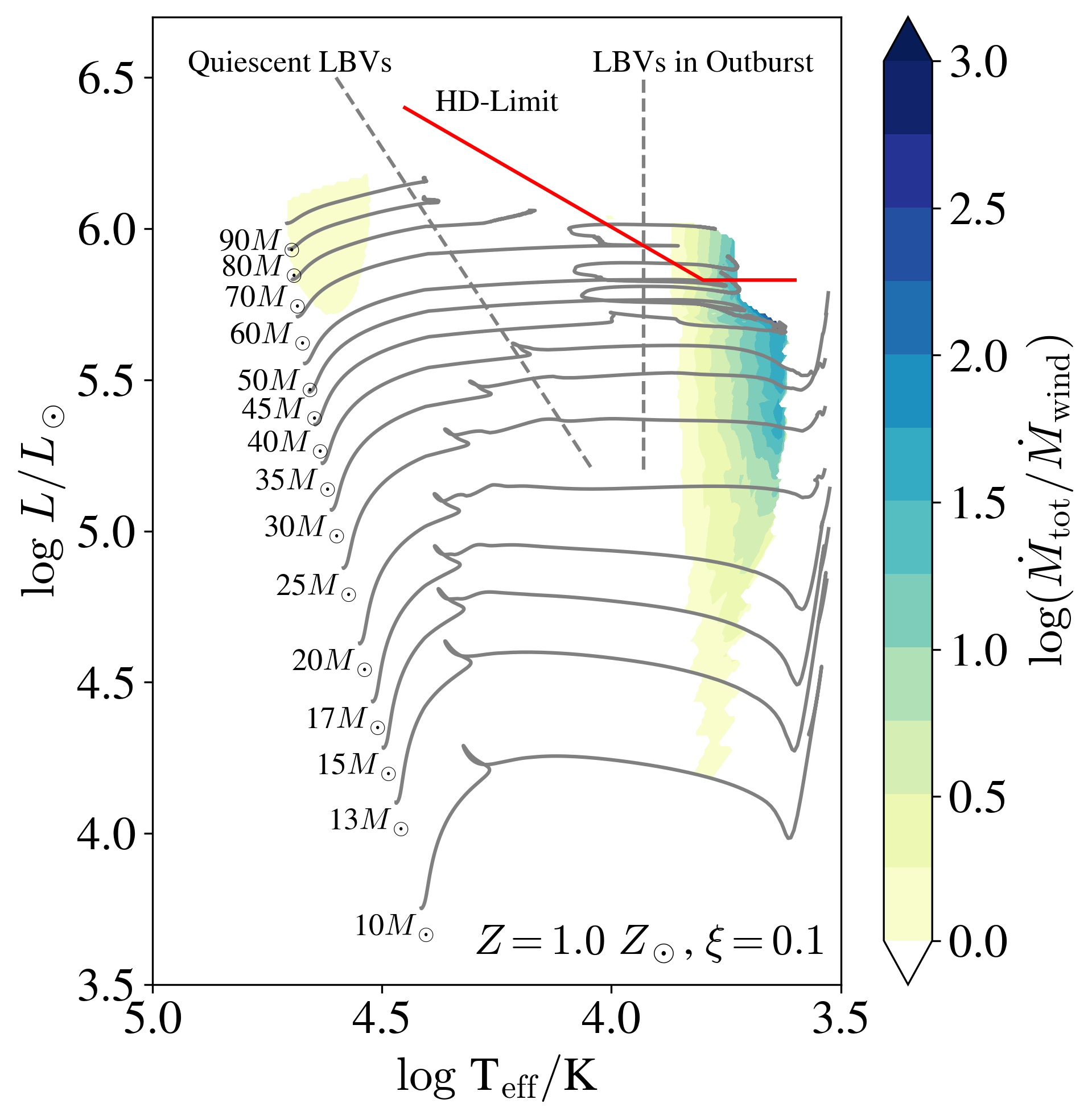}{0.5\textwidth}{}
          \fig{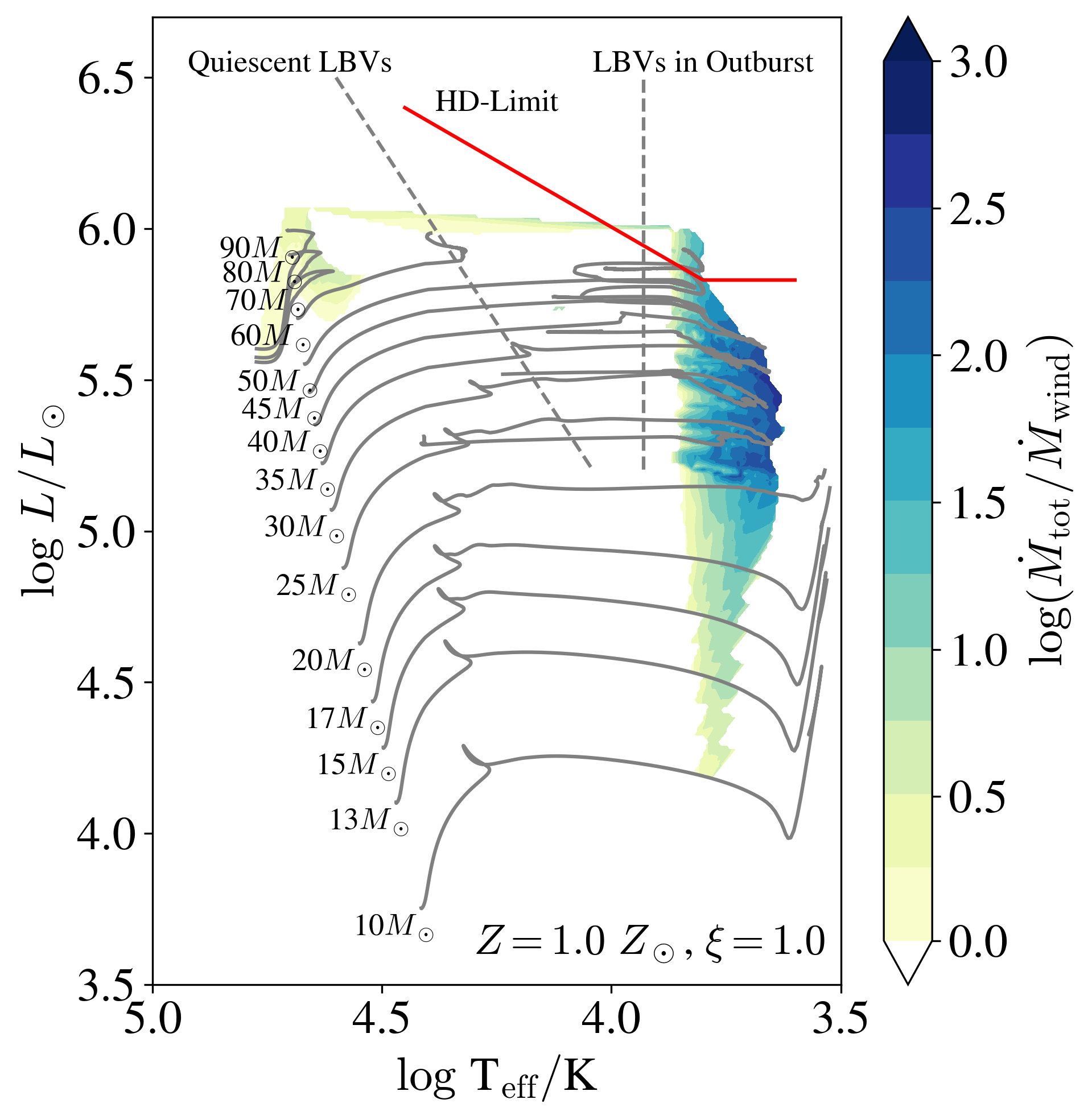}{0.5\textwidth}{}
          }
    \gridline{\fig{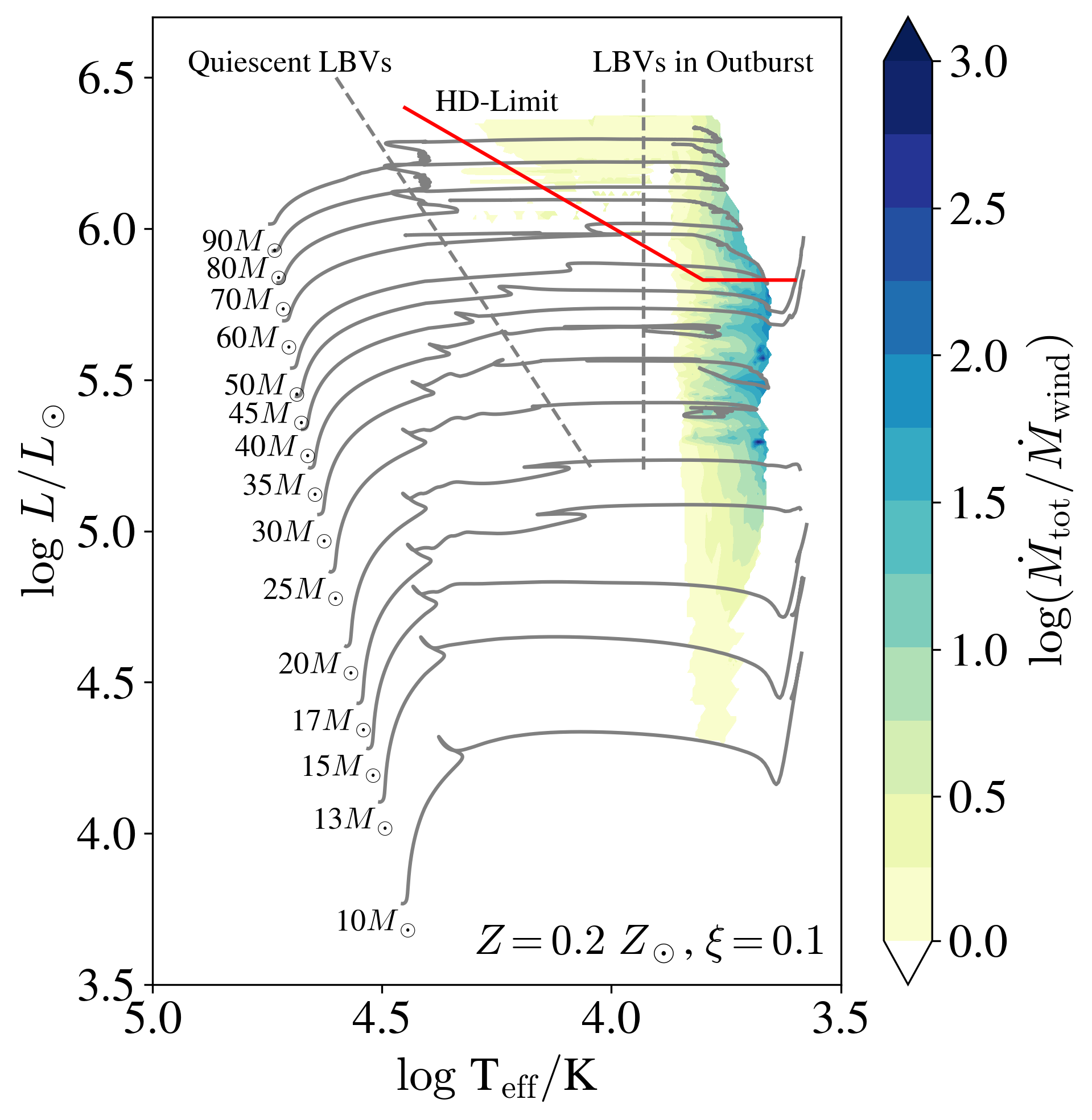}{0.5\textwidth}{}
          \fig{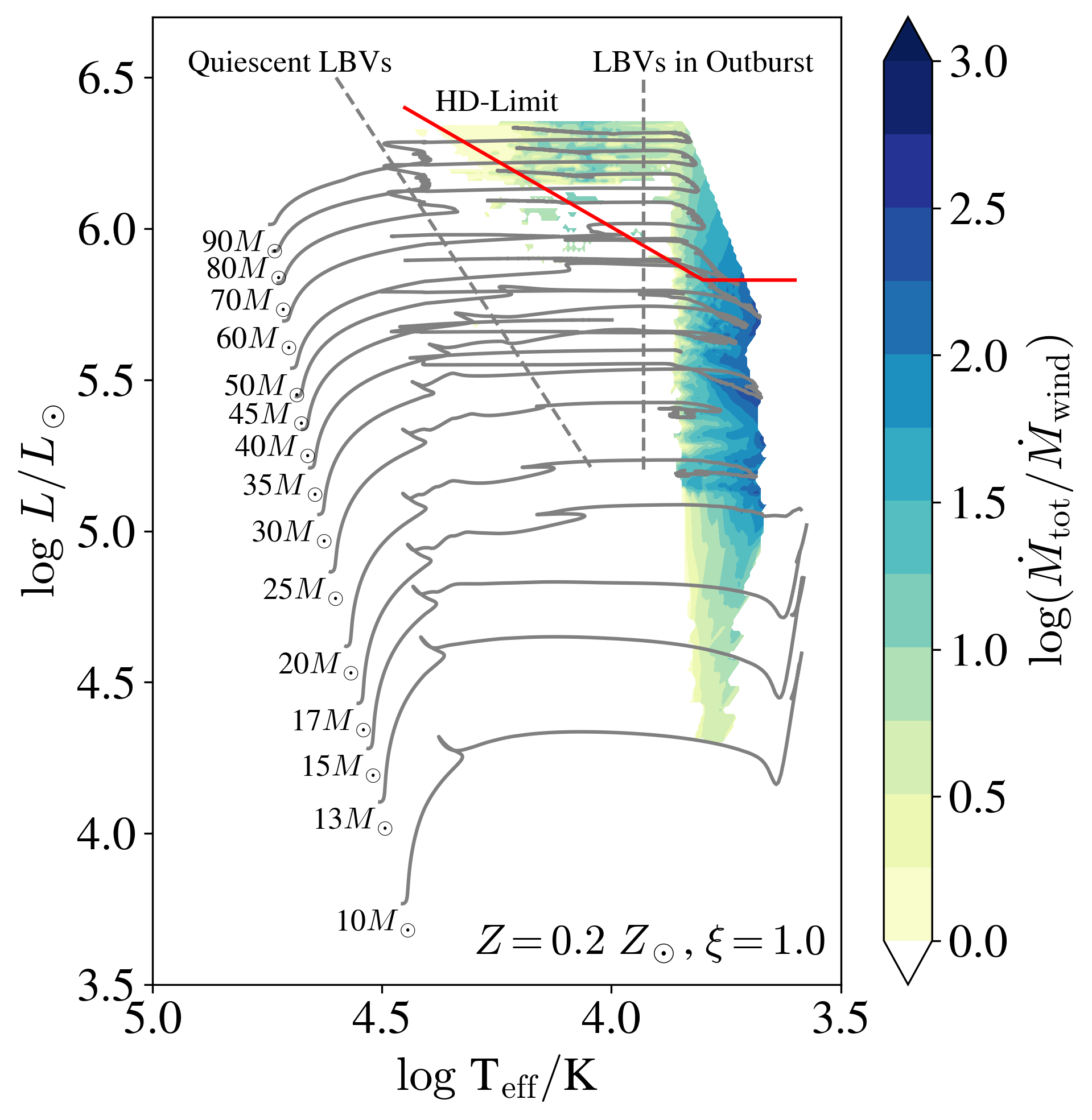}{0.5\textwidth}{}
          }
    \caption{Ratio of total mass loss against wind mass loss rates across the H-R diagram. Left column: $\xi=0.1$. Right column: $\xi=1.0$. Top row: $Z=1~Z_\odot$. Bottom row: $Z=Z_\mathrm{SMC}=0.2~Z_\odot$. Colored contours here show the ratio between total mass loss and wind mass loss, with legend on the right. Cooler colors (blues) indicate that eruptive mass loss is more dominant over stellar wind. All axes, labels, and other lines follow Figure~\ref{fig:res}.}\label{fig:ratiores}
\end{figure*}

Additionally, for stars with initial masses greater than $50~M_\odot$, some mass loss of order $10^{-5}~M_\odot\mathrm{yr}^{-1}$ at higher $\log(T_\mathrm{eff}/\mathrm{K}) > 4.5~\mathrm{K}$ is predicted by our eruptive mass loss model (seen in top left of Figure~\ref{fig:res}). This is due to the iron opacity peak in the stellar envelope at $\sim 1.8\times 10^5$~K (\citealt{paxton:13,Jiang:15}, see Figure~\ref{fig:opacity_peaks}), and follows the results from 3D models in \citet{Jiang:15}. However, mass loss in the $\log(T_\mathrm{eff}/\mathrm{K}) > 4.5$ region is dominated by stellar winds rather than eruptive mass loss. This can be seen in Figure~\ref{fig:ratiores}, which shows the fractional increase in overall mass loss due to eruptive mass loss (total mass loss/wind mass loss). Regions in the H-R diagram in which eruptive mass loss dominates substantially over stellar winds are marked in blue, which only occurs at lower temperatures. Therefore, from Figure~\ref{fig:ratiores}, we determine that our eruptive mass loss is dominant over stellar winds in the $3.5 < \log(T_\mathrm{eff}/\mathrm{K}) < 4.0$ vertical band region.

Similarly, as shown in the bottom panels in Figures~\ref{fig:res}, \ref{fig:trackres}, and \ref{fig:ratiores}, a lower metallicity of $Z=0.2~Z_\odot\approx Z_\mathrm{SMC}$ also predicts a lack of red supergiants above $\sim 20~M_\odot$ at $\xi = 1.0$. Since lower metallicities lead to lower opacities, increasing $L_\mathrm{Edd}$ and thereby reducing $\Gamma_\mathrm{Edd, rad} = L_\mathrm{rad}/L_\mathrm{Edd}$, some of the numerical challenges involving locally super-Eddington regions in the star (see Section~\ref{sec:massive_stars}) are avoided and the high-mass $Z=0.2~Z_\odot\approx Z_\mathrm{SMC}$ models are able to evolve further than those with $Z=1~Z_\odot$. The lower panels of Figure~\ref{fig:res} shows that at both metallicities, stars evolved with $\xi=1.0$ exhibit higher rates of eruptive mass loss of up to $\sim10^{-2}M_\odot~\mathrm{year}^{-1}$ whereas those evolved with $\xi=0.1$ feature correspondingly lower mass loss rates of up to $\sim10^{-3}M_\odot~\mathrm{year}^{-1}$. Unsurprisingly, the region above the Humphreys-Davidson (HD) limit exhibits high levels of eruptive mass loss, as this observational limit corresponds to regions on the H-R diagram where surface stellar luminosities transition to become super-Eddington. Additionally, this region above the HD limit also exceeds the de Jager limit \citep{deJager:1984}, wherein supersonic turbulent motions in the photosphere may have a destabilizing effect. Models at $Z=1~Z_\odot$ exhibited higher maximum eruptive mass loss rates by up to an order of magnitude than those at $Z=0.2~Z_\odot$ for $M<40~M_\odot$. For $M>40~M_\odot$, the $Z=0.2~Z_\odot$ models were able to evolve further and therefore exhibited higher maximum eruptive mass loss rates by at least an order of magnitude compared to $Z=1~Z_\odot$ models. The maximum eruptive mass loss rate across all models is $\approx 3\times10^{-2}~M_\odot~\mathrm{year}^{-1}$ (occurring for $\xi=1.0$), which is consistent with observations (see Section~\ref{sec:intro}).

\vspace{2cm}

\section{Discussion and Conclusions} \label{sec:discussion}

We studied eruptive mass loss in massive stars of varying masses (between $10-100~M_\odot$) at two different metallicities ($Z_\mathrm{SMC} \approx 0.2~Z_\odot $ and $1~Z_\odot$) with an implementation of an energy-based eruptive mass loss model in the stellar evolution code \texttt{MESA}. Our model's mass loss rate is determined by considering photon diffusion through regions in the outer stellar envelope of these massive stars. As parts of the outer envelope may be locally super-Eddington, the excess (above Eddington) energy output from radiation may act on the fluid mass faster than dynamical, convective, and acoustic processes and exceed the binding energy of the overlaying layers, leading to mass loss.

We find that, at $\xi=1.0$ (a model efficiency of one for the eruptive mass loss) and at both metallicity choices, stars with initial masses greater than $20~M_\odot$ no longer evolve to become red supergiants. This suggests that eruptive mass loss could be a physical mechanism that contributes to the observed lack of RSGs above $20-30~M_\odot$, despite recent careful work determining relatively low mass-loss rates in well-characterized isolated RSGs \citep{Beasor:20,Beasor:2023,Decin:2024}.

Alternative explanations other than eruptive mass loss for the lack of RSGs above $20-30~M_\odot$ include envelope inflation \citep[e.g.,][]{Sanyal:15, Sanyal:17} and binary interactions \citep[e.g.,][]{Smith:2014, Morozova:2015, Margutti:2017}. In the case of envelope inflation, \citet{Sanyal:15} ran 1D hydrodynamic stellar evolution models on core hydrogen burning massive stars and found that stars with initial mass $M>40~M_\odot$ locally exceeded the Eddington limit in the partial ionisation zones of iron, helium, or hydrogen. The stars featured density inversions in their envelopes and experienced an increase of radius of up to a factor of $\sim40$, which alters the temperature and density structure such that envelope opacity decreases until the Eddington limit is no longer exceeded. While the authors found that the inflated envelopes remained bound, they noted that their models may have overestimated convective energy transport due to their implementation of MLT (therefore reducing $L_\mathrm{rad}$) and that multi-dimensional hydrodynamical simulations are needed to fully determine the stability of density inversions in stellar envelopes. Additionally, the authors speculatively mentioned that eruptions are a possible consequence of their very coolest models. Alternatively, binary interactions such as envelope stripping can lead to the formation of Wolf-Rayet stars that may never enter the RSG phase \citep[e.g.,][]{Dray:2003, Meynet:2011, Grafener:2012}. However, mass-transferring binaries leave behind accretors and mergers, many of which are expected to evolve to the RSG phase \citep[e.g.,][]{Podsiadlowski:93,Aldering:94,Maund:04,Claeys:11,Yoon:17,Schneider:20,Renzo:21, Renzo:23}.

Additionally, our work may offer a possible solution to the ``Blue Supergiant Problem'' wherein many blue supergiants (BSGs) having been observed despite the expectation of rapid evolution through the BSG phase. BSGs are hot, luminous stars with luminosities up to $\log(L/L_\odot) \approx 6.0$ and temperatures up to $\log(T_\mathrm{eff})/\mathrm{K} \approx 4.4$ \citep[e.g., HD 5980 by][]{Pickering:01, Cannon:18, Koenigsberger:10}. It is generally thought that BSGs are either due to changes in stellar structure via mixing during evolution across the Hertzsprung gap \citep[e.g.,][]{Iben:93,Sugimoto:00,deMink:09,deMink:13,Schootemeijer:19}, or post-main-sequence interactions such as mass transfer or mergers \citep[e.g.,][]{Podsiadlowski:92,Braun:95,Justham:14,deMink:14}. The eruptive mass loss presented in this work can be regarded as a possible intrinsic post-main-sequence behavior that may contribute to the formation of BSGs.

The energy-based model for eruptive mass loss presented in this work draws parallels with the ``geyser model'' presented in \citet{Maeder:1992}.  In this model, the mass loss rate during a shell ejection event in a massive star is estimated by assuming the mass above super-Eddington peaks is ejected and accounting for the inward-movement of these peaks at the local thermal timescale. Our model does not explicitly account for this effect since it would operate at timescales shorter than our code timestep. As such, the ``geyser model'' may further enhance eruptive mass loss, though our model may approximately account for an average of such effects.

We emphasize that due to the numerical challenges of modelling massive stars in 1D (see Section~\ref{sec:massive_stars}), our models were only evolved far enough into core helium burning to probe the rightmost section of the HR diagram where red supergiants are found. Nevertheless, our work demonstrates that a physically-motivated energy-based mass loss model can adequately describe eruptive mass loss in massive stars. Additionally, we note that our work here treated the mass loss model's efficiency factor $\xi$ as a free parameter and explored the implications of $\xi=1.0$ and $\xi=0.1$. An observational study aimed at tuning this efficiency factor using LBV mass loss observations will be conducted in future work.

\section*{Acknowledgments}
SJC and CC acknowledge support from NSF grant AST-131576. The Flatiron Institute is supported by the Simons Foundation. 

\newpage
\appendix{}\label{appendix}

\begin{figure}[ht!]
        \begin{center}
            \includegraphics[width=1.0\columnwidth]{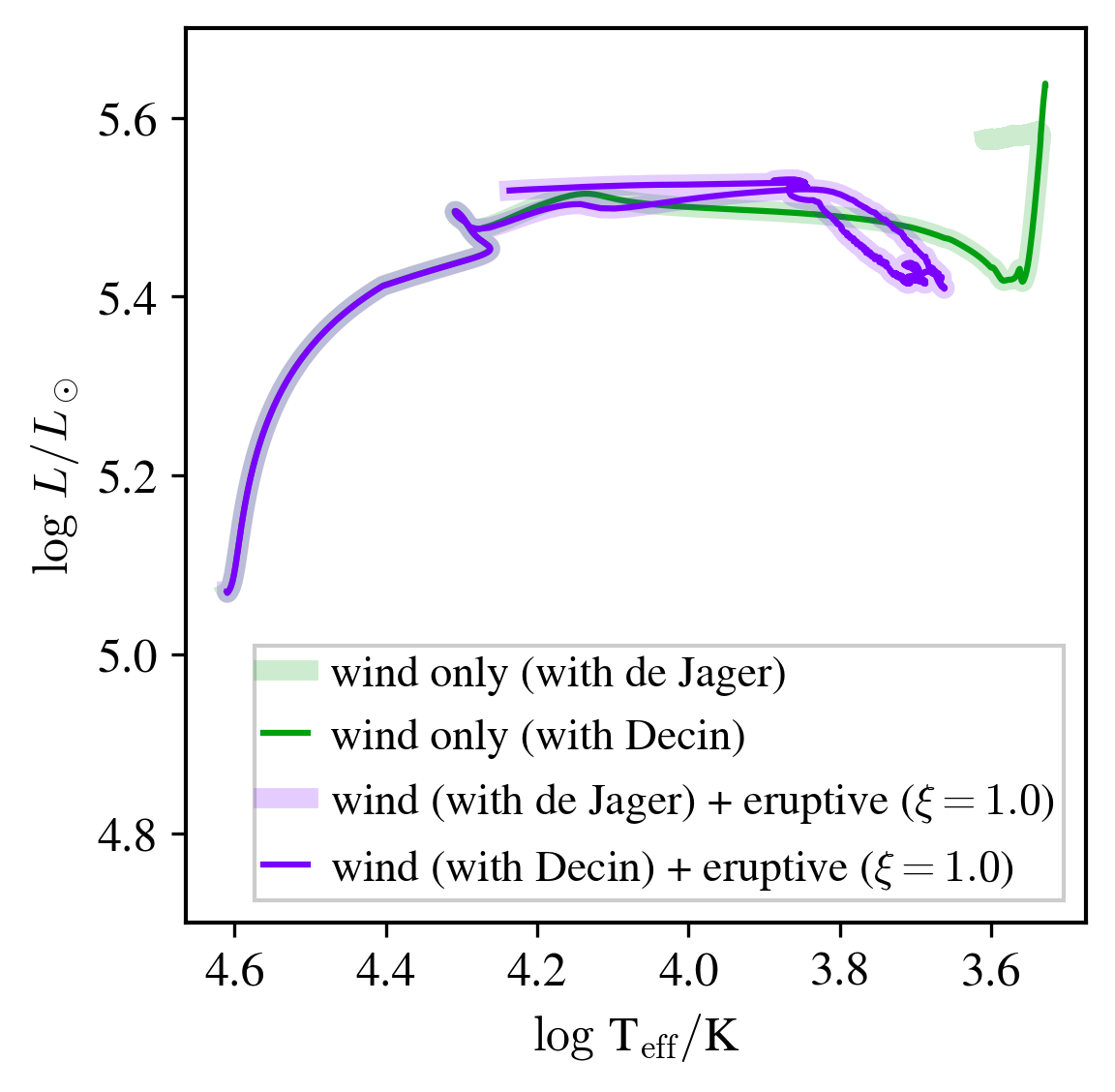}
        \end{center}
        
        \caption{Evolutionary tracks of $M_\mathrm{ZAMS}=30~M_\odot$, $Z=1~Z_\odot$ stars. H-R tracks for wind only (green) and eruptive mass loss $+$ wind (purple) models. The models marked with ``Decin" contain the modification described in Section~\ref{sec:methods}, while those using the old \citet{deJager:88} rates are marked with ``de Jager''. From the differing green and purple tracks, eruptive mass loss prevents evolution to RSG phase.}\label{fig:tracks_alt}
\end{figure}

For RSG winds, we choose to adopt \citet{Decin:2024} rather than one of the schemes included in \texttt{MESA}'s  \texttt{Dutch} prescription \citep[e.g.,][]{deJager:88, Nieuwenhuijzen:90} for which mass loss rates are overestimated (see Section~\ref{sec:intro}). While doing so alters stellar evolution in the RSG phase, it does not affect our eruptive mass loss result: the left panel of Figure~\ref{fig:tracks_alt} compares models run with \citet{Decin:2024} and with \citet{deJager:88} mass loss for $T_\mathrm{eff} < 4000~\mathrm{K}$. The \citet{Decin:2024} winds only affects the very late RSG phase, and does not impact the model with eruptive wind in the case shown (in purple) since such models do not evolve cool enough to enter the \citet{Decin:2024} regime. We note that although \citet{Decin:2024} produces mass loss rates calibrated to careful observations in well-characterized individual stellar populations (in broad agreement with \citet{Beasor:2023}'s correction to the \citet{Beasor:20} rates), it contains in it a dependence on the initial mass of stars as estimated from cluster characteristics, which need not correspond to the initial mass of a stellar model in \texttt{MESA}. A more proper treatment correcting for this when comparing models to stellar populations might include a variable scaling factor akin to the \verb|dutch_scaling_factor| for the \texttt{\textquotesingle{Dutch}\textquotesingle} stellar wind prescription.

\newpage
\bibliography{references}{}
\bibliographystyle{aasjournal}

\end{document}